\documentclass[12pt]{article}
\usepackage{graphicx}
\usepackage{epsfig}
\usepackage{epstopdf}
\DeclareGraphicsExtensions{.pdf,.eps,.png,.jpg,.mps}
\usepackage{url}
\usepackage[implicit=false]{hyperref}
\usepackage{cite}

\def\lsim{\mathrel{\rlap{\lower3pt\hbox{\hskip0pt$\sim$}}
     \raise1pt\hbox{$<$}}}         
\def\gsim{\mathrel{\rlap{\lower4pt\hbox{\hskip1pt$\sim$}}
     \raise1pt\hbox{$>$}}}         

\usepackage{amsmath}
\usepackage{amsfonts}

\begin{document}
\begin{titlepage}

\centerline{\Large \bf Cryptoasset Factor Models}
\medskip

\centerline{Zura Kakushadze$^\S$$^\dag$\footnote{\, Zura Kakushadze, Ph.D., is the President and CEO of Quantigic$^\circledR$ Solutions LLC,
and a Full Professor at Free University of Tbilisi. Email: zura@quantigic.com}}
\bigskip

\centerline{\em $^\S$ Quantigic$^\circledR$ Solutions LLC}
\centerline{\em 1127 High Ridge Road \#135, Stamford, CT 06905\,\,\footnote{\, DISCLAIMER: This address is used by the corresponding author for no
purpose other than to indicate his professional affiliation as is customary in
publications. In particular, the contents of this paper
are not intended as an investment, legal, tax or any other such advice,
and in no way represent views of Quantigic$^\circledR$ Solutions LLC,
the website \url{www.quantigic.com} or any of their other affiliates.
}}
\centerline{\em $^\dag$ Free University of Tbilisi, Business School \& School of Physics}
\centerline{\em 240, David Agmashenebeli Alley, Tbilisi, 0159, Georgia}
\medskip
\centerline{(September 6, 2018)}

\bigskip
\medskip

\begin{abstract}
{}We propose factor models for the cross-section of daily cryptoasset returns and provide source code for data downloads, computing risk factors and backtesting them out-of-sample. In ``cryptoassets" we include all cryptocurrencies and a host of various other digital assets (coins and tokens) for which exchange market data is available. Based on our empirical analysis, we identify the leading factor that appears to strongly contribute into daily cryptoasset returns. Our results suggest that cross-sectional statistical arbitrage trading may be possible for cryptoassets subject to efficient executions and shorting.
\end{abstract}
\medskip

\end{titlepage}

\newpage

\section{Introduction}

{}Crytoassets\footnote{\, By ``cryptoassets" here we mean digital cryptography-based assets such as cryptocurrencies (e.g., Bitcoin), as well as the plethora of various other digital ``coins" and ``tokens" (minable as well as non-minable) that have arisen in the recent years. For our specific purposes here, all digital assets that have data on \url{https://coinmarketcap.com} a priori are included in ``cryptoassets".} have a sizable (albeit highly volatile) total market capitalization measuring in hundreds of billions of dollars. Superfluously, there is also a sizable number of these cryptoassets, edging toward 2,000 as of this writing. The question we ask -- and, at least to a some degree, answer -- in this note is this: {\em Are there common (risk) factors underlying the cross-section of cryptoasset returns?}

{}There are no evident ``fundamentals" for cryptoassets based on which one could attempt to build ``fundamental" long-horizon factors for cryptoassets akin to value, growth, etc., for stocks.\footnote{\, For some literature on long-horizon factors for equities, see, e.g., \cite{Amihud2002}, \cite{Ang2006}, \cite{Anson2013}, \cite{Asness1995}, \cite{Asness2001}, \cite{Asness2000}, \cite{Banz1981}, \cite{Basu1977}, \cite{Carhart1997}, \cite{Fama1992}, \cite{Fama1993}, \cite{Fama1996}, \cite{Haugen1995}, \cite{Jegadeesh1993}, \cite{Lakonishok1994}, \cite{Liew2000}, \cite{Pastor2003}, \cite{Scholes1977}.} However, even for stocks, on short horizons (e.g., overnight returns) in some sense things become simpler as the longer-horizon ``fundamentals" (again, such as value and growth) are no longer relevant \cite{CvSRM}. This underlies the construction of the 4-factor model of \cite{4FM} for equity returns on short-horizons. It is therefore natural to extend the ideas set forth in \cite{4FM} to cryptoassets, to wit, to daily open-to-close returns.

{} And this is precisely what this note does. We consider 4(+) factors, to wit, cap (or size, based on market cap), mom (momentum), hlv (based on average intraday volatility), and vol (or liquidity, based on average daily dollar volume). By running out-of-sample Fama-MacBeth regressions \cite{FMacB} and computing annualized t-statistic from the time series of the corresponding regression coefficients, we conclude that vol is not a good predictor (with a possible exception of the previous day's volume). One possible explanation is that cryptoassets on average trade much less in comparison to their market caps (low ``turnover"), so vol does not actually meaningfully measure liquidity. The other three factors cap, mom and hlv do add value, with mom leading by a large margin. In fact, momentum from the day before the previous day is also predictive. The sign of the mom regression coefficient is negative, which indicates a mean-reversion effect in cryptoasset daily returns.\footnote{\, To our knowledge, our analysis here is the first of its kind. For some cryptoasset investment and trading related literature, see, e.g.,  \cite{Alessandretti2018}, \cite{Amjad2017}, \cite{Baek2014}, \cite{Bariviera2017}, \cite{Bouoiyour2016}, \cite{Bouri2017}, \cite{Brandvold2015}, \cite{Briere2015}, \cite{Cheah2015}, \cite{CheungRS2015}, \cite{Ciaian2015}, \cite{Colianni2015}, \cite{Donier2015}, \cite{Dyhrberg2015}, \cite{Eisl2015}, \cite{Gajardo2018}, \cite{Garcia2015}, \cite{Georgoula2015}, \cite{Harvey2016}, \cite{JiangLiang2017}, \cite{KimKim2016}, \cite{Kristoufek2015}, \cite{Chuen2018}, \cite{Li2018}, \cite{LiewLB2018}, \cite{Nakano2018}, \cite{Ortisi2016}, \cite{Shah2014}, \cite{VanAlstyne2014}, \cite{WangVergne2017}.}

{}The remainder of this note is organized as follows. In Section \ref{sec2} we describe the data, define our factors, and discuss the results of our regressions. Section \ref{sec3} briefly concludes with some comments. Appendix \ref{app.A} gives R source code for data downloads and running factor regressions.\footnote{\, The source code given in Appendix \ref{app.A} hereof is not written to be ``fancy" or optimized for speed or in any other way. Its sole purpose is to illustrate the algorithms described in the main text in a simple-to-understand fashion. Some important legalese is relegated to Appendix \ref{app.B}.} Tables and figures summarize our results.

\section{Factors}\label{sec2}
\subsection{Setup and Data}

{}Unlike stocks, barring any special circumstances such as unexpected halts in trading, cryptoassets trade continuously, 24/7. So, while there are notions of ``open" and ``close" for cryptoassets, their meanings are different from those for stocks. For our purposes here, again, barring any special circumstances, ``open" on any given day means the price right after midnight (UTC time), while ``close" on any given day means the price right before midnight (UTC time). In this regard, absent trading halts, the open on a given day is very close to the close of the previous day. The high and low prices then have the usual meaning within the 24 hour window between the open and the close. And the volume is the dollar volume traded in said 24 hour interval. All the prices are also measured in dollars, as is the market cap. We use index $i=1,\dots,N$ to label $N$ different cryptoassets cross-sectionally, and index $s=0,1,2,\dots$ to denote the dates, with $s=0$ corresponding to the most recent date in the time series. So: $P^C_{is}$ (or, equivalently, $P^C_{i,s}$) is the close price for the cryptoasset labeled by $i$ on the day labeled by $s$; $P^O_{is}$ is the open price; $P^H_{is}$ is the high price; $P^L_{is}$ is the low price; $V_{is}$ is the dollar volume; $C_{is}$ is the market cap. All our data was freely downloaded from \url{https://coinmarketcap.com} (see below).

{}Next, we define our daily returns as open-to-close intraday returns:
\begin{equation}
 R_{is} = \ln\left({P^{C}_{is} / P^{O}_{is}}\right)
\end{equation}
The use of the log-return (or ``continuously compounded" return) is intentional here. For small values it is approximately the same as the standard (``single-period") return defined as
\begin{equation}
 {\widetilde R}_{is} = P^{C}_{is} / P^{O}_{is} - 1
\end{equation}
However, cryptoassets can be very volatile, on average, much more so than stocks, and log-returns ``smooth out" the outliers somewhat, so below we use $R_{is}$.

{}Unlike with stocks, there are no ``dividends" to worry about for cryptoassets, to wit, in terms of adjusting prices for dividends. However, an issuer can split its cryptoasset. Thus, Xaurum had a forward split 8000-to-1 on August 23, 2016, so its price decreased accordingly. Unfortunately, \url{https://coinmarketcap.com} does not adjust historical prices for splits, and there does not appear to be a simple source to look up historical splits data. Fortunately, for the purposes of analyzing our factor models here, such splits are immaterial as all our factors are defined such that they are unaffected by splits. Note that market cap and dollar volume are unaffected, only prices are. However, in our factor definitions for any given day we only use ratios of intraday prices, which therefore are also unaffected. In this paper, the only place where splits become important is when we plot price weighted indexes, and we account for the aforesaid Xaurum split there (see below).

\subsection{Factor Model}

{}The factor model is of the form
\begin{equation}\label{reg}
 R_{is} = \sum_{A = 1}^K \beta_{iAs}~ f_{As} + \varepsilon_{is}
\end{equation}
Here $K$ is the number of risk factors, $f_{As}$ are the $K$ factor returns, $\varepsilon_{is}$ are the residuals, and $\beta_{iAs}$ are the factor loadings. We include the intercept in $\beta_{iAs}$, so for a given date $s$, the $N\times K$ matrix $\beta_{iAs}$ contains a column equal the unit $N$-vector, which we will take to be the first column in $\beta_{iAs}$. Below, instead of using the index $A$, we will denote each column by $\beta^{\rm{\scriptstyle{\{name\}}}}_{is}$, where \{name\} stands for the name by which we refer to the corresponding factor: ``int" for the intercept, ``cap" for size, ``mom" for momentum, ``hlv" for intraday high and low based volatility, and ``vol" for volume. These are direct analogs of the short-horizon factors for stocks defined in \cite{4FM}. We also consider another binary factor (dummy variable), which we refer to as ``mnbl", based on whether the cryptoasset is minable.

{}To determine whether a given factor adds value, as in \cite{FMacB}, we use annualized t-statistic $\tau_A$ for each factor, which can be computed using the daily returns $f_{As}$ as follows:
\begin{eqnarray}
 &&\tau_A = \sqrt{365}~{\overline{f}_A \over \sigma_A}\\
 &&\overline{f}_A = {1\over T} \sum_{r=1}^T f_{A,t + r}\\
 &&\sigma_A^2 = {1\over {T-1}} \sum_{r=1}^T \left[f_{A,t + r} - {\overline f}_A\right]^2
\end{eqnarray}
Here $t$ is the beginning of the period for which the t-statistic is computed, and $T$ is the length of said period. Throughout, all time quantities are measured in days. Also, the annualization factor\footnote{Note that t-statistic is a horizon-dependent quantity; it scales as the square root of the horizon.} is $\sqrt{365}$ (as opposed to $\sqrt{252}$ for stocks) as cryptoassets trade 24/7. The meaning of $\tau_A$ is that it is the annualized Sharpe ratio \cite{Sharpe1994} based on the time series of the corresponding factor returns $f_{As}$ (without subtracting any ``benchmark" return from the mean return $\overline{f}_A$). So, in the ``zeroth approximation", $\tau_A$ above 3 would indicate that the corresponding factor is a relevant predictor, and $\tau_A$ below 2 would indicate that it is a poor predictor.

\subsubsection{Intercept}\label{sub.beta}

{}The intercept plays the role of the ``market beta":
\begin{equation}
 \beta^{\rm{\scriptstyle{int}}}_{is} \equiv 1
\end{equation}
Neutrality w.r.t. $\beta^{\rm{\scriptstyle{int}}}_{is}$ is dollar neutrality, which is approximate market neutrality.

\subsubsection{Cap (Size)}

{}Size is the natural logarithm of the market cap\footnote{\, In \cite{4FM} for stock returns the log of the price was used instead as the market cap is the price times the shares outstanding, and the latter change negligibly on short horizons. The same generally holds for cryptoassets, so we could use the price instead of the market cap. However, considering the split adjustment issues mentioned above, it is simpler to use market cap.}
\begin{equation}\label{size}
 \beta^{\rm{\scriptstyle{cap}}}_{is} = \ln\left(C_{i, s+1}\right)
\end{equation}
So, on date $s$ we use the previous day's market cap, which is 100\% out-of-sample.

\subsubsection{Momentum}

{}There are various ways to define the momentum factor loading. For our purposes here, we will define it as the previous day's open-to-close return:
\begin{equation}\label{mom}
 \beta^{\rm{\scriptstyle{mom}}}_{is} = R_{i,s+1}
\end{equation}
Again, this definition is 100\% out-of-sample. Below we will also analyze momenta (which we refer to as mom1, mom2, \dots) defined using the days prior to the previous day:
\begin{eqnarray}\label{mom+}
 &&\beta^{\rm{\scriptstyle{mom1}}}_{is} = R_{i,s+2}\\
 &&\beta^{\rm{\scriptstyle{mom2}}}_{is} = R_{i,s+3}\\
 &&\dots
\end{eqnarray}

\subsubsection{Intraday Volatility}

{}There are various ways to define the intraday volatility factor. Below we will use the following simple definition:\footnote{\, For alternative definitions, see \cite{4FM}.}
\begin{eqnarray}\label{hlv}
 &&\beta^{\rm{\scriptstyle{hlv}}}_{is} =  {1\over 2}~\ln\left(U_{is}\right)\\
 &&U_{is} = {1\over d_{hlv}} \sum_{r=1}^{d_{hlv}} \left({{P^{H}_{i, s+r} - P^L_{i, s+r}}\over P^{C}_{i, s+r}}\right)^2
\end{eqnarray}
Averaging over the previous $d_{hlv}$ days (which is done 100\% out-of-sample) is necessary to smooth out the noise. Since here we are dealing with a variance-like quantity (as opposed to a correlation-like quantity), looking back $d_{hlv}$ days does not introduce out-of-sample instabilities associated with correlations and time series based betas. Below we use the values $d_{hlv}=20, 15, 10,5$. Also, note that (\ref{hlv}) is the logarithmic intraday volatility. This is because, as is generally the case with volatility, the intraday volatility itself (without taking the log) has a skewed, long-tailed distribution for higher-end values, and, among other things, as a factor it would adversely interfere with the intercept and also result in unnaturally skewed regression residuals $\varepsilon_{is}$.

\subsubsection{Volume}\label{sub.vol}

{}For stocks, the average daily dollar volume is a measure of liquidity. For cryptoassets the situation is murkier (see below). Nonetheless, following \cite{4FM} we define the volume factor loading as follows:
\begin{eqnarray}\label{vol}
 && \beta^{\rm{\scriptstyle{vol}}}_{is} = \ln\left({\overline V}_{is}\right)\\
 && {\overline V}_{is} = {1 \over d_{vol}} \sum_{r=1}^{d_{vol}} V_{i, s+r}
\end{eqnarray}
Recall that $V_{is}$ is the daily dollar volume, so ${\overline V}_{is}$ is the average daily dollar volume over the $d_{vol}$ days immediately preceding the day labeled by $s$ (so it is 100\% out-of-sample). Below we use the values $d_{vol} = 20, 15, 10, 5, 3, 1$. For $d_{vol}=1$ we have the previous day's dollar volume. The reason for including it will become clear below.

\subsubsection{``Is-Minable" Dummy}\label{sub.mnbl}

{}Some cryptoassets are minable and some are not. Therefore, it is natural to consider a binary factor loading defined as follows (notwithstanding that a priori it is not clear why it would have an impact on daily returns):
\begin{eqnarray}\label{mnbl}
 \beta^{\rm{\scriptstyle{mnbl}}}_{is} = \begin{cases}
          1  &  \mbox{ if minable}  \\
          0  &  \mbox{ if not minable}
        \end{cases}
\end{eqnarray}
This factor loading is independent of the time index $s$ and only depends on $i$.

\subsection{Estimation Period and Universe}\label{sub.univ}

{}We downloaded\footnote{\, R source code for data downloads and running factor regressions is given in Appendix \ref{app.A}.} the data from \url{https://coinmarketcap.com} for all cryptoassets as of August 19, 2018 (so the most recent date in the data is August 18, 2018), whose number was 1,855. Out of those, 1,851 had downloadable data, albeit for many various fields were populated with ``?", which we converted into NAs. Despite a seemingly large number of cryptoassets, the useful data going back any reasonable number of days only exists for a fraction of them. In order to be able to run our regressions (see below), we only kept cryptoassets with non-NA price (open, close, high, low), volume and market cap data, with an additional filter that no null volume was allowed either (to avoid contamination by stale prices). There were only 362 cryptoassets with such properties starting August 18, 2018 (inclusive) and going back $365+20+1$ days (i.e., 1 year ``padded" with additional 21 days to be able to compute 20-day moving averages out-of-sample), 129 cryptoassets when the filters were applied to $2\times 365+20+1$ days (i.e., 2 years ``padded" with additional 21 days), and only 66 cryptoassets when the filters were applied to $3\times 365+20+1$ days (i.e., 3 years ``padded" with additional 21 days).\footnote{\, Actually, 2 cryptoassets had apparently ``artifact" stale prices in the second and third year (looking back), so they had to be excluded from the corresponding regressions (see below).} So, the cross-section is pretty ``thin", nowhere near as rich as for stocks. However, we must work with what we have.

\subsection{Regressions and Results}\label{sub.res}

{}So, we run regressions of the daily returns $R_{is}$ over the loadings $\beta_{iAs}$ (computed as above) for various periods and their subperiods described above and in Tables \ref{table1} through \ref{table18}. Tables \ref{table1} through \ref{table10} include int (intercept), cap (size), mom (momentum), hlv (intraday volatility) and vol (volume), for various values of $d_{hlv}$ and $d_{vol}$. From these tables it appears that the average daily dollar volume for $d_{vol}>1$ is a poor predictor. For $d_{vol}=1$ (i.e., the previous day's dollar volume) the results are at best mixed and certainly not ground-shaking. While it is possible that the previous day's volume adds value, it might do so more efficiently as an additional filter (when, e.g., defining a trading signal) as opposed to a stand-alone factor. Tables \ref{table11} and \ref{table12}, which exclude vol and include only int, cap, mom and hlv, appear to support this conclusion. Further, hlv appears to be rather stable w.r.t. the choice of $d_{hlv}$. The mom factor leads by a large margin in all periods (see Tables \ref{table11} and \ref{table12}). The cap and hlv factors appear to work well except in the first year (meaning, going {\em back} in time from August 18, 2018, so this is the most recent 1-year period) for the smaller universes based on 129 and 66 cryptoassets (see above). However, these factors work much better for the same period for the larger universe based on 362 cryptoassets (see above), which subsumes the aforesaid smaller universes. Therefore, the most likely explanation would appear to be that this is due to the smallness of the cross-sectional samples for these smaller universes. Further, the hlv factor appears to be the weakest among cap, mom and hlv. However, removing hlv worsens the results (see Table \ref{table13}), as does removing the intercept (see Table \ref{table14}).\footnote{\, The intercept has variable t-statistic across various periods and universes, including changing its sign. However, this is not surprising as the intercept plays the role of the ``market beta", which is expected to be highly variable on general grounds, as it is for stocks as well as other assets.}

{}Considering how strong the mom factor is, it is natural to also look at momenta from days prior to the previous day, i.e., mom1, mom2, \dots (defined above). Tables \ref{table15} and \ref{table16} suggest that mom1 indeed appears to be a good predictor. Going beyond mom1 (see Table \ref{table17}, which also includes mom2, mom3 and mom4) unsurprisingly gives mixed results as any effect from momentum is expected to decay with time. Note that the regression coefficients for the momenta are negative, which suggests a substantial {\em mean-reversion} effect in the aforesaid cryptoasset returns. Including the dummy variable mnbl defined above does not appear to add value (see Table \ref{table18}).

\subsection{``Sanity Check"}

\begin{figure}[ht]
\centering
\includegraphics[scale=1.0]{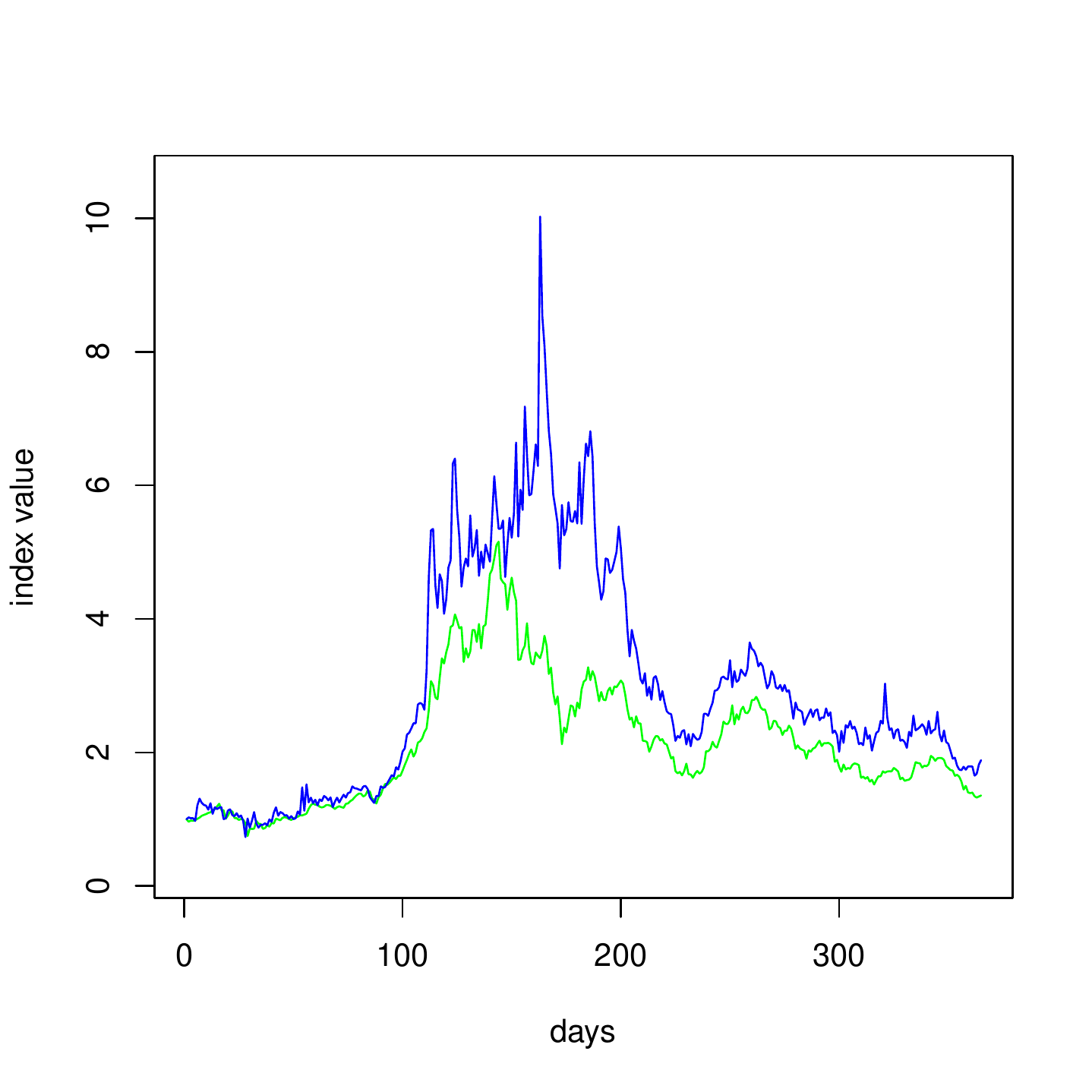}
\caption{The market cap weighted index (bottom) and price weighted index (top) for the same 1-year period and cryptoasset universe as in Table \ref{table1}. Both indexes are normalized to 1 on the first day of the period. The spike in the price weighted index is due to a high-priced cryptoasset (42-coin) with a small market cap.}
\label{Figure1}
\end{figure}

\begin{figure}[ht]
\centering
\includegraphics[scale=1.0]{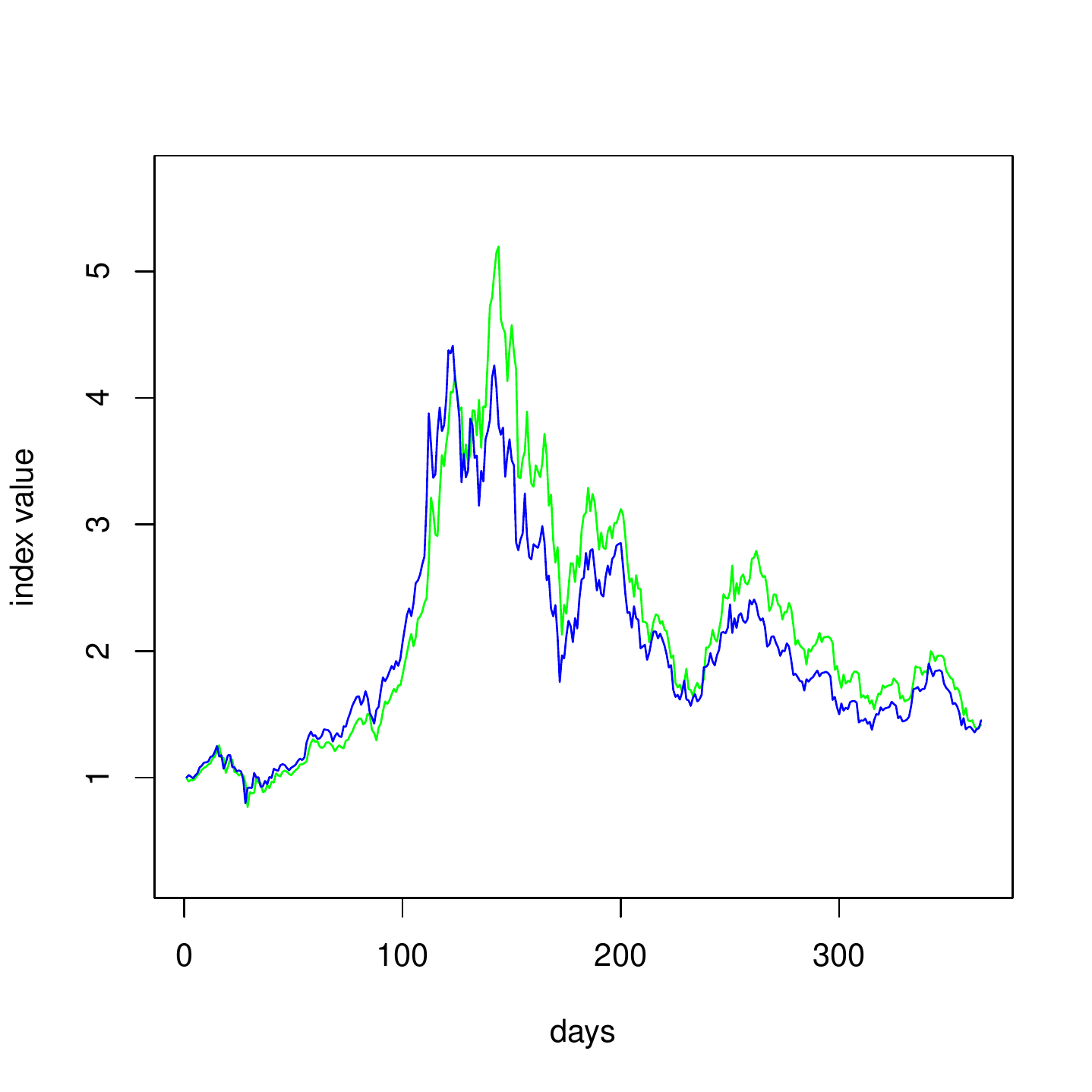}
\caption{The market cap weighted index (top) and price weighted index (bottom) for the same 1-year period and cryptoasset universe as in Table \ref{table3}. Both indexes are normalized to 1 on the first day of the period.}
\label{Figure2}
\end{figure}

\begin{figure}[ht]
\centering
\includegraphics[scale=1.0]{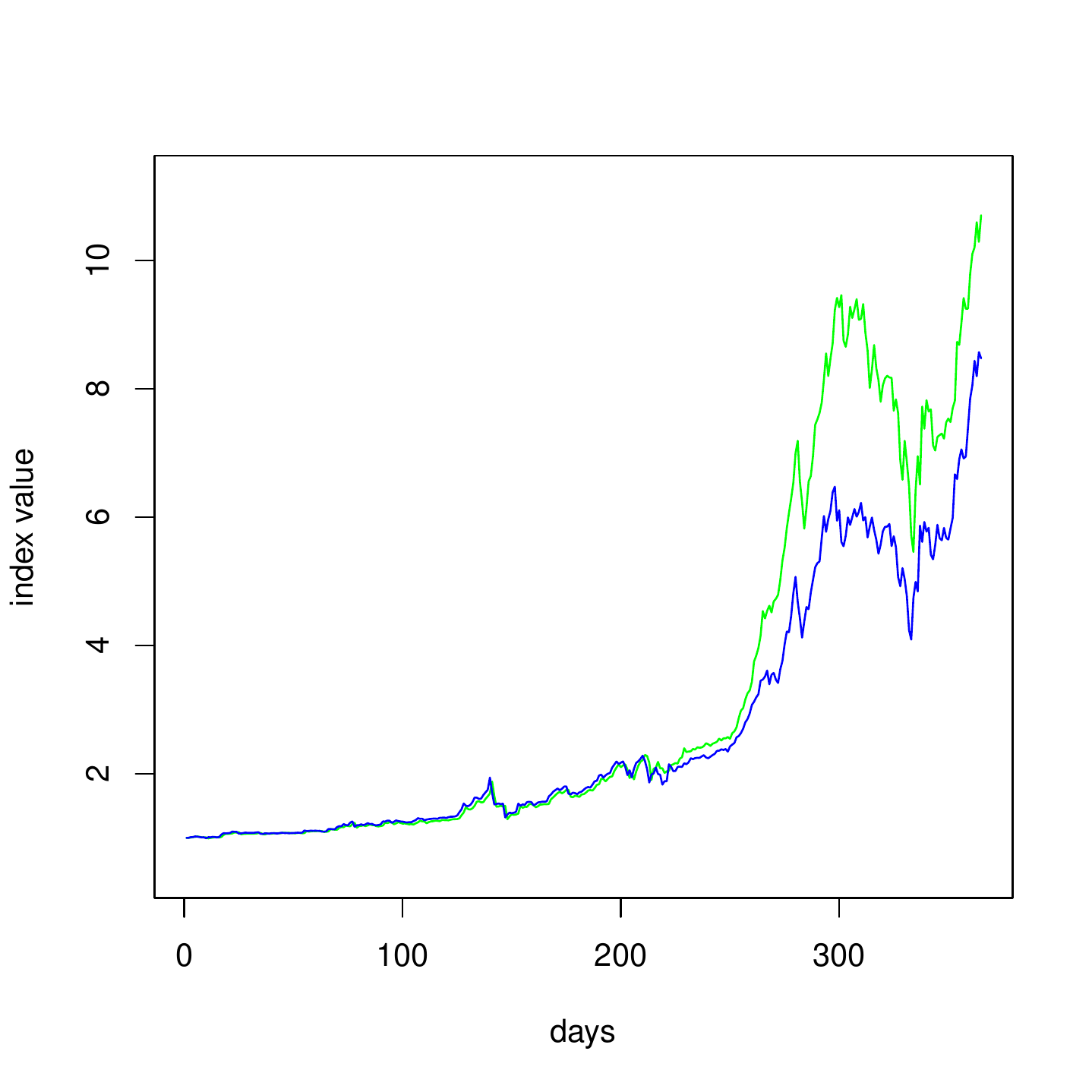}
\caption{The market cap weighted index (top) and price weighted index (bottom) for the same 1-year period and cryptoasset universe as in Table \ref{table4}. Both indexes are normalized to 1 on the first day of the period.}
\label{Figure3}
\end{figure}

\begin{figure}[ht]
\centering
\includegraphics[scale=1.0]{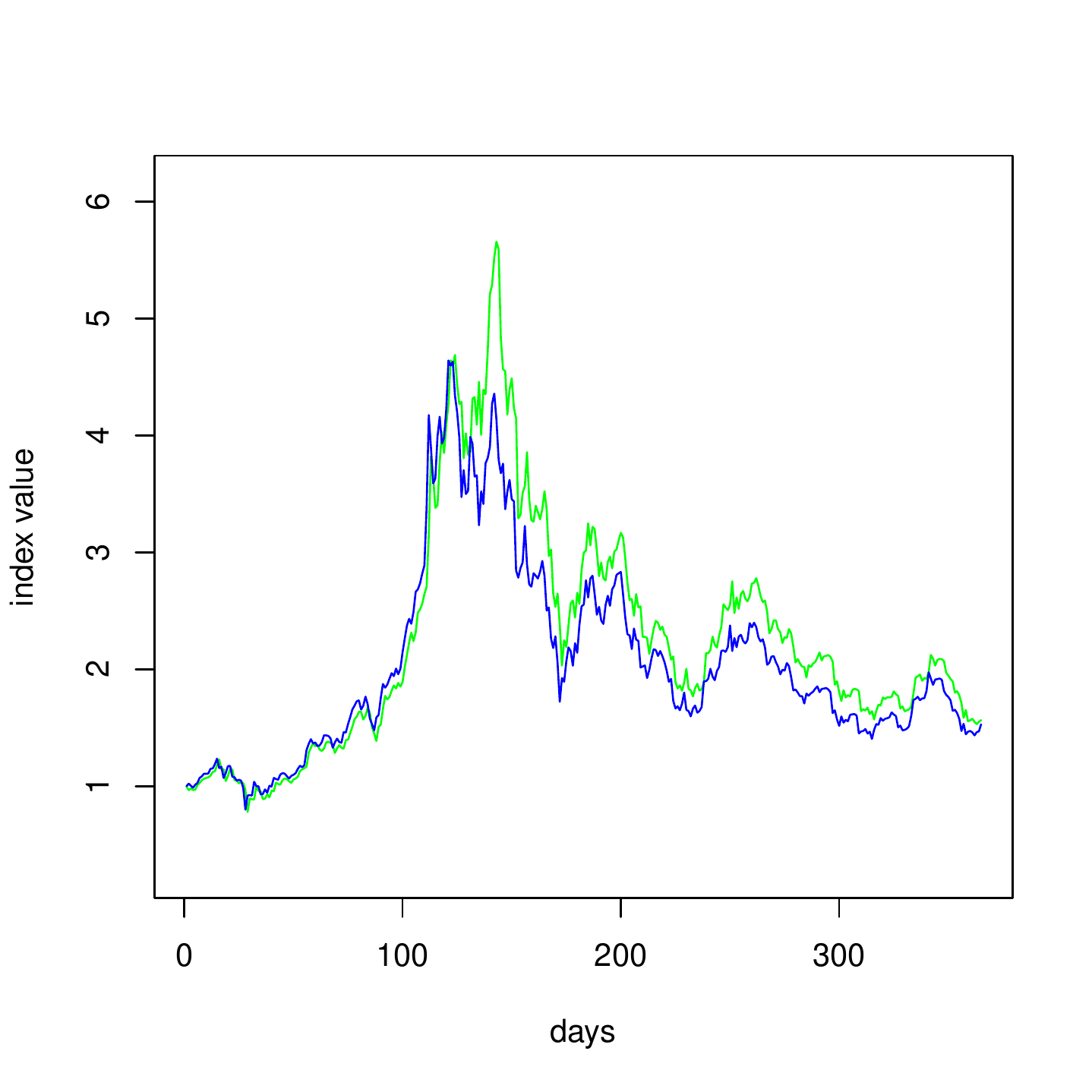}
\caption{The market cap weighted index (top) and price weighted index (bottom) for the same 1-year period and cryptoasset universe as in Table \ref{table8}. Both indexes are normalized to 1 on the first day of the period.}
\label{Figure4}
\end{figure}

\begin{figure}[ht]
\centering
\includegraphics[scale=1.0]{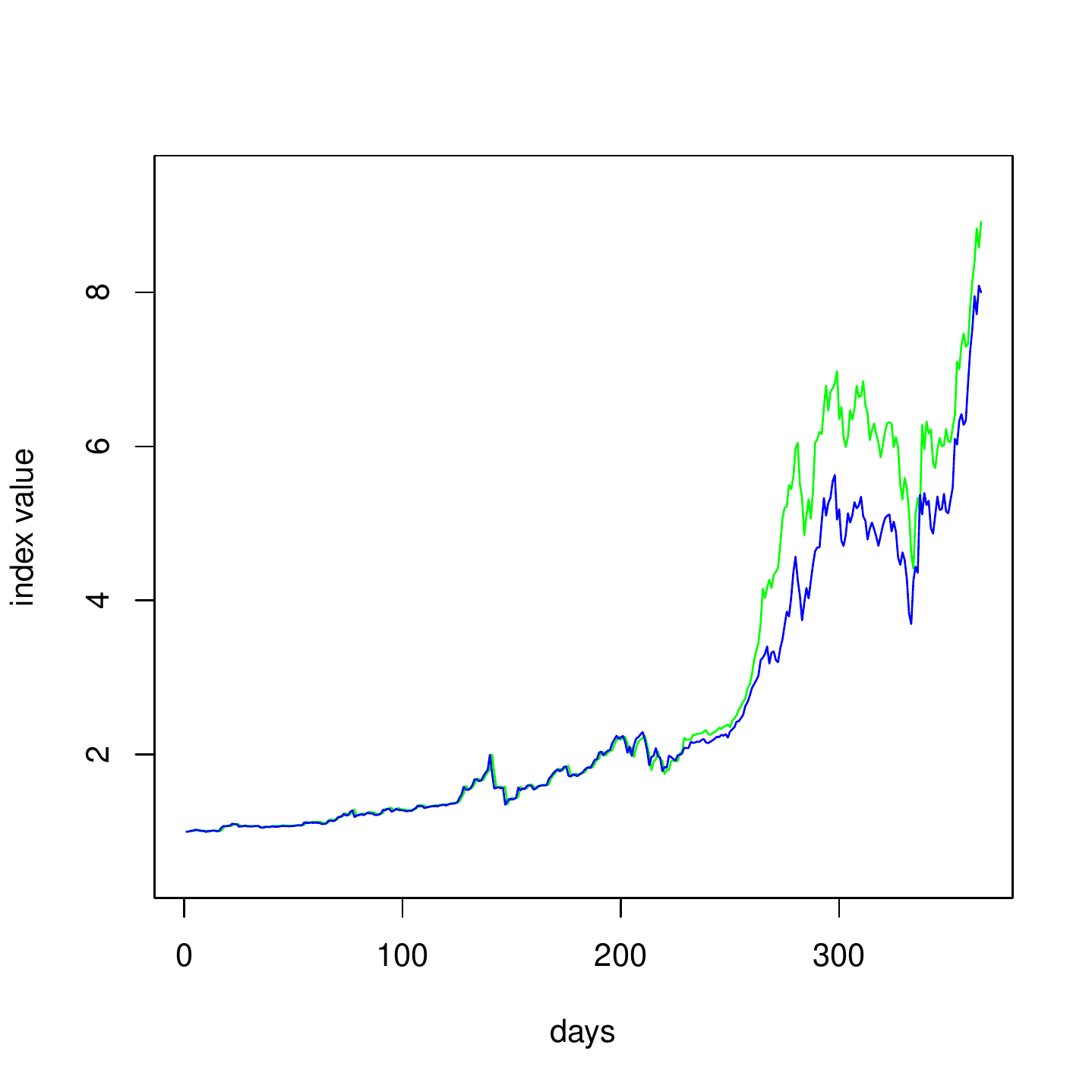}
\caption{The market cap weighted index (top) and price weighted index (bottom) for the same 1-year period and cryptoasset universe as in Table \ref{table9}. Both indexes are normalized to 1 on the first day of the period.}
\label{Figure5}
\end{figure}

\begin{figure}[ht]
\centering
\includegraphics[scale=1.0]{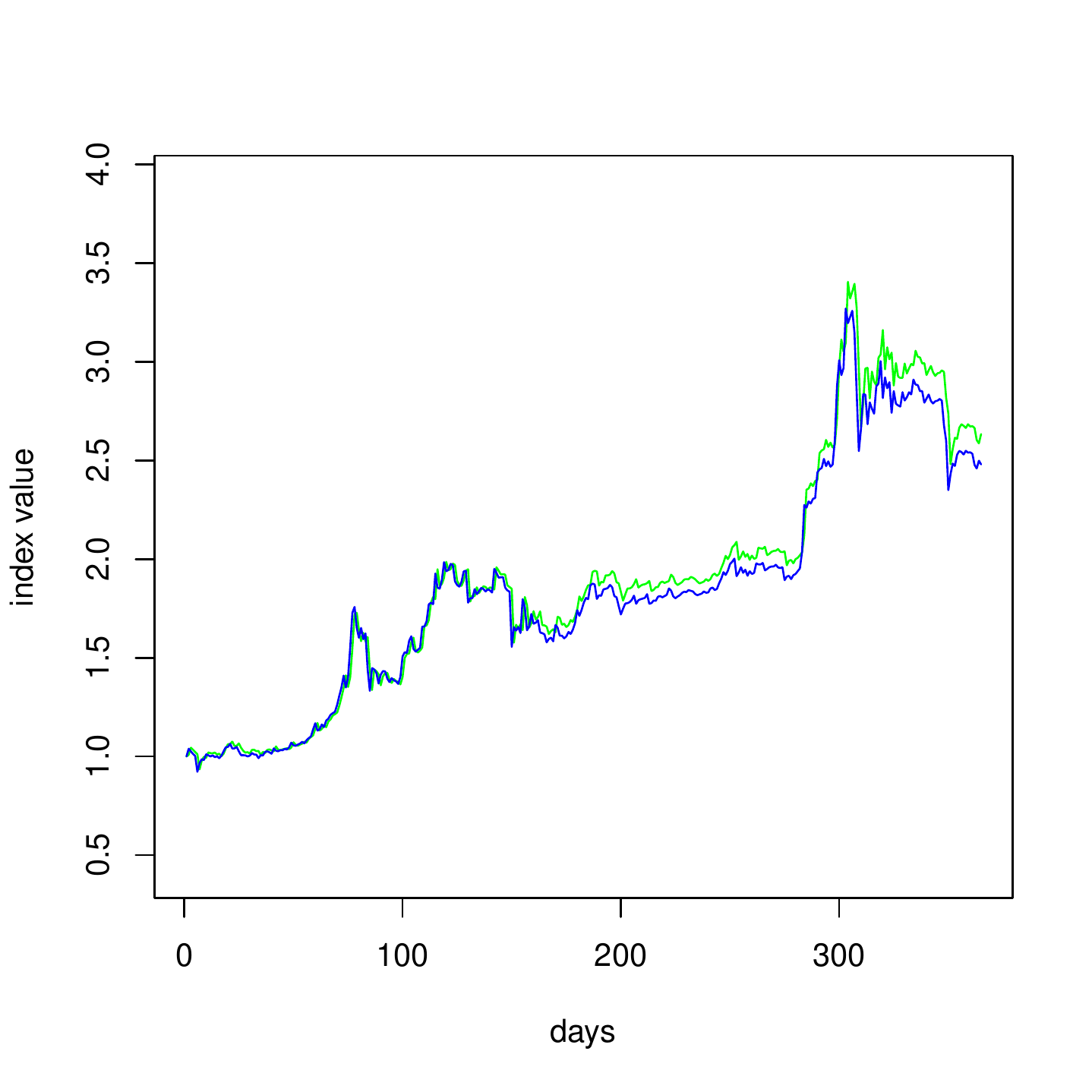}
\caption{The market cap weighted index (top) and price weighted index (bottom) for the same 1-year period and cryptoasset universe as in Table \ref{table10}. Both indexes are normalized to 1 on the first day of the period.}
\label{Figure6}
\end{figure}

{}Considering the ``flukes" in the performance of cap and hlv during the most recent 1-year period for smaller universes (see above), it makes sense to get at least a superficial visual confirmation that nothing utterly ``odd" happened with those universes during that period. A simple thing to do here is to look at the performance of ``market indexes" built based on the aforesaid 362, 129 and 66 cryptoasset universes. Thus, we can construct the following indexes (among myriad others):
\begin{eqnarray}
 &&I^{cap}_s = \gamma_{cap} \sum_{i=1}^N C_{is}\\
 &&I^{prc}_s = \gamma_{prc} \sum_{i=1}^N {\widetilde P}^C_{is}
\end{eqnarray}
Here $I^{cap}_s$ is a market cap weighted index for the universe of $N$ cryptoassets, and the normalization coefficient $\gamma_{cap}$ is fixed such that $I^{cap}_s$ equals 1 for the value of $s$ corresponding to the earliest date in the time period for which $I^{cap}_s$ is constructed. The price weighted index $I^{prc}_s$ is constructed similarly, except that the prices ${\widetilde P}^C_{is}$ are obtained from the closing prices $P^C_{is}$ by adjusting for splits. In our case we only had to adjust for the Xaurum split mentioned above. The results are plotted in Figures \ref{Figure1} through \ref{Figure6}. There appears to be nothing ``peculiar" about these market indexes for the most recent 1-year period and the aforesaid smaller universes, except for the spike in the price weighted index in Figure \ref{Figure1}. However, this spike is due to a single cryptoasset called 42-coin, which is highly priced with a small market cap. Incidently, this is a good example of why price weighted indexes can be misleading.

\section{Concluding Remarks}\label{sec3}

\begin{figure}[ht]
\centering
\includegraphics[scale=0.65]{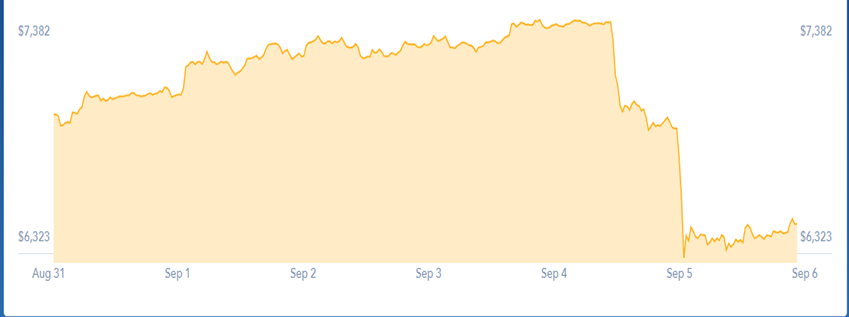}
\caption{The Bitcoin price drop (along with the broad cryptoasset market) on September 5, 2018. This snapshot was generated using an interactive chart on \url{https://www.coinbase.com}.}
\label{Figure7}
\end{figure}

{}Despite the superfluous ubiquity of cryptoassets, the amount of cross-sectionally available data is still limited as for most cryptoassets it simply does not go far back enough. Cryptoassets come and go, many coins and tokens disappear a short while after issuance, and many are perceived as being scams to raise a quick buck and run with the unsuspecting or uninformed investors' money. This is still a very young field despite Bitcoin having being around for over 9 years, so the kind of analyses performed in \cite{4FM} on 2,000, 3,000 and even 4,000 stock tickers going back 5 years is simply impossible for cryptoassets at this nascent stage in their development. It might take another 5-10 years to collect that kind of data, depending on their survival rate -- and assuming the whole field does not disappear due to regulatory or some other (less foreseeable) issues.\footnote{\, Thus, as this note was being finalized, the cryptoasset market had yet another crash on September 5, 2018 on the news that Goldman Sachs reportedly is putting on hold its plans for a cryptocurrency trading desk \cite{Campbell2018}. Also see Figure \ref{Figure7}.} Only time will tell.

{}Nonetheless, it is pleasant to observe that the 3 short-horizon factors discussed in \cite{4FM} for stocks, to wit, cap (size), mom (momentum) and hlv (intraday volatility) work well for cryptoassets as well. The fact that vol (volume) does not seem to add value for cryptoassets (at least beyond the previous day's volume) is not necessarily surprising as volume is not directional. In fact, on its own it is not a good predictor for stocks either and only works when combined with the other factors. One possible explanation for why vol does not add value for  cryptoassets is that it is less evident how well vol describes actual liquidity in cryptoasset markets considering a very different (from stocks) fee structure and executions. Thus, one way to quantify and see the difference between cryptoassets and stocks (as it relates to volume) is to compare their respective cross-sections of ratios of, say, the 20-day average daily dollar volume to the market cap, which is a measure of average daily ``turnover". For the 362 cryptoasset universe (see above) as of August 18, 2018, the cross-sectional summary of this ratio is as follows: Min = $2.00 \times 10^{-10}$, 1st Quartile = $6.87 \times 10^{-7}$, Median = $3.03 \times 10^{-6}$, Mean = $3.16 \times 10^{-5}$, 3rd Quartile = $1.44 \times 10^{-5}$, Max = $2.71 \times 10^{-3}$. For stocks, for the universe of the 500 highest market cap stocks as of (a randomly chosen date) August 31, 2010, that summary reads: Min = $1.05 \times 10^{-5}$, 1st Quartile = $6.09 \times 10^{-3}$, Median = $8.57 \times 10^{-3}$, Mean = $1.05 \times 10^{-2}$, 3rd Quartile = $1.25 \times 10^{-2}$, Max = $1.08 \times 10^{-1}$. The bottom line is that on relative basis (i.e., based on the average daily ``turnover") cryptoassets do not trade anywhere near as much as stocks, so it is not that surprising that the volume is not a good predictor for cryptoassets.

{}The fact that momentum dominates as a factor for cryptoasset returns means that on short horizons the market is strongly mean-reverting (cross-sectionally). This in turn implies that if one could short a large cross-section of cryptoassets and trade them on short horizons cost-effectively, one might be able to trade a cross-sectional dollar-neutral mean-reversion statistical arbitrage strategy with cryptoassets, along the lines of similar strategies for stocks -- perhaps one day soon.

\appendix
\section{R Source Code: Downloads and Regressions}\label{app.A}

{}In this appendix we give R (R Project for Statistical Computing, \url{https://www.r-project.org/}) source code for downloading cryptoasset data and running factor regressions discussed in the main text. The code is straightforward and self-explanatory. The first function is {\tt{\small crypto.data()}}, which downloads the names, symbols, market caps and various other data for all cryptoassets available at any given time from \url{https://coinmarketcap.com} and outputs said data in the file {\tt{\small crypto.cap.txt}}. Among other things, in the column labeled {\tt{\small URL}}, this file contains strings that are used on \url{https://coinmarketcap.com} for historical data downloads, which are not the same as symbols (or names). In fact, oddly, there can be cases where two different cryptoassets have the same symbol, so symbols are not unique differentiators. The function {\tt{\small crypto.data()}} internally calls three auxiliary functions {\tt{\small shared.get.webpage()}}, {\tt{\small shared.parse.html()}} and {\tt{\small shared.write.table()}}. These auxiliary functions are given in \cite{SIC} (which is freely downloadable) with the names {\tt{\small sec.get.webpage()}}, {\tt{\small sec.parse.html()}}, {\tt{\small sec.write.table()}}.

{}The second function {\tt{\small crypto.hist.prc(date)}} (which also uses the aforesaid 3 auxiliary functions) downloads data for all cryptoassets in the file {\tt{\small crypto.cap.txt}}. There can be a few names for which the data is bad, which are skipped and the corresponding fields from the 11th column (labeled {\tt{\small URL}}) of the file {\tt{\small crypto.cap.txt}} are written in the file {\tt{\small crypto.bad.txt}}. The historical data for the rest of the names is written into tab-delimited text files (one file per name) in the directory {\tt{\small CryptoHistData}}. The sole input {\tt{\small date}} (which is in the format {\tt{\small yyyymmdd}}) of the function {\tt{\small crypto.hist.prc(date)}} corresponds to the date as of which to download the data. This can be modified, e.g., to include the specific lookback date through which to do download the historical data. As-is, this function downloads all available (on \url{https://coinmarketcap.com}) historical data starting with {\tt{\small date}} and going back in time (so the lookbacks are name-dependent).

{}The third function {\tt{\small crypto.prc.files()}} aggregates the data stored in the individual files in the directory {\tt{\small CryptoHistData}} by data type (close, open, high, low, volume, market cap). It reads the file {\tt{\small crypto.bad.txt}} to skip the names therein and generates tab-delimited text files with $N$ rows and $d$ columns, where $N$ is the number of all names excepting those in {\tt{\small crypto.bad.txt}}, and $d$ is the number of days for which there is data for the first name in {\tt{\small crypto.cap.txt}}, which is the oldest cryptoasset Bitcoin (as the names in {\tt{\small crypto.cap.txt}} are ranked by market cap, and Bitcoin has the largest one). These files are {\tt{\small cr.prc.txt}} (close price), {\tt{\small cr.open.txt}} (open price), {\tt{\small cr.high.txt}} (high price), {\tt{\small cr.low.txt}} (low price), {\tt{\small cr.vol.txt}} (dollar volume), and {\tt{\small cr.cap.txt}} (market cap). The function also generates two single-column text files {\tt{\small cr.name.txt}} (names of the cryptoassets in the same order as all the other files), and {\tt{\small cr.mnbl.txt}} (1 if the name is minable, otherwise 0). All missing data (including all fields with ``?") is populated with NAs.

{}The fourth and last function {\tt{\small crypto.prc()}} reads the aforesaid aggregated data files, constructs factor loadings internally referred to as {\tt{\small av}} (average daily dollar volume), {\tt{\small size}} (market cap), {\tt{\small mom}} (momentum) and its delayed variations {\tt{\small mom1}} through {\tt{\small mom4}}, {\tt{\small hlv}} (intraday volatility), as well as {\tt{\small mnbl}} (the ``is-minable" dummy), and runs regressions, which include {\tt{\small int}} (intercept) and are used for computing the annualized t-statistic based on the time series of regression coefficients. The inputs of {\tt{\small crypto.prc()}} are {\tt{\small days}} (the length of the selection period used in fixing the cryptoasset universe by applying the aforesaid non-NA data and non-zero volume filters, which period is further ``padded" -- see below), {\tt{\small back}} (the length of the skip period, i.e., how many days to skip in the selection period before the lookback period), {\tt{\small lookback}} (the length of the lookback period over which the regression is run), {\tt{\small d.r}} (the extra ``padding" added to the selection period plus one day, so the moving averages can be computed out-of-sample; we take {\tt{\small d.r = 20}}), {\tt{\small d.v}} (the same as $d_{vol}$ defined in Section \ref{sec2}), and {\tt{\small d.i}} (the same as $d_{hlv}$ defined in Section \ref{sec2}). Thus, for instance, in Table \ref{table9} we have {\tt{\small days = 3 * 365}} (so with {\tt{\small d.r = 20}} the ``padded" selection period contains $1116 = 3\times 365 + 20 + 1$ days), {\tt{\small back = 365}}, and {\tt{\small lookback = 365}}. Therefore, the regressions are run over 365 days by skipping the most recent 365 days, starting with the 366th day in the time series and going back in time (i.e., over the 2nd year in the 3-year time series, looking {\em back}). The function {\tt{\small crypto.prc()}} also constructs and plots market cap and price weighted indexes based on the universes and periods over which the regressions are run.\\
\\
{\tt{\small
\noindent crypto.data <- function ()\\
\{\\
\indent require(XML)\\
\indent require(httr)\\
\\
\indent url <- "https://coinmarketcap.com/all/views/all/"\\
\indent z <- x <- shared.get.webpage(url)\\
\indent x <- shared.parse.html(x, keyword = "table")\\
\indent u <- c(x[22:28])\\
\indent for(i in 1:length(u))\\
\indent \{\\
\indent \indent u1 <- grep(u[i], x)\\
\indent \indent x <- x[-u1]\\
\indent \}\\
\indent u1 <- grep("[*]", x)\\
\indent x1 <- x[-u1]\\
\indent hdr <- c("Rank", "Name", "Symbol", "MktCap", "Price",\\
\indent \indent \indent "Supply", "Volume", "Ch1h", "Ch24h", "Ch7d",\\
\indent \indent \indent "URL", "Minable")\\
\indent x1 <- x1[-(1:10)]\\
\indent x1 <- matrix(x1, length(x1)/11, 11, byrow = T)\\
\indent x1 <- x1[, -2]\\
\indent x1 <- gsub(",", "", x1)\\
\indent x1 <- gsub("\textbackslash\textbackslash\$", "", x1)\\
\indent x1 <- gsub("\%", "", x1)\\
\indent x1 <- cbind(x1, rep("", nrow(x1)), rep("Y", nrow(x1)))\\
\indent y <- grep("currency-symbol visible-xs", z)\\
\indent z1 <- z[y]\\
\indent y <- grep("link-secondary", z1)\\
\indent z1 <- z1[y]\\
\indent y <- grep("href", z1)\\
\indent z1 <- z1[y]\\
\indent y <- grep("currencies", z1)\\
\indent z1 <- z1[y]\\
\indent if(length(z1) != nrow(x1))\\
\indent \indent stop("ERROR")\\
\indent x <- x[-(1:10)]\\
\indent for(i in 1:length(z1))\\
\indent \{\\
\indent \indent u <- unlist(strsplit(z1[i], "/"))[3]\\
\indent \indent x1[i, 11] <- u\\
\indent \indent u1 <- grep("[*]", x[1:11])\\
\indent \indent if(length(u1) > 0)\\
\indent \indent \{\\
\indent \indent \indent x1[i, 12] <- "N"\\
\indent \indent \indent x <- x[-(1:12)]\\
\indent \indent \}\\
\indent \indent else\\
\indent \indent \indent x <- x[-(1:11)]\\
\indent \}\\
\\
\indent x1 <- rbind(hdr, x1)\\
\indent shared.write.table(x1, "crypto.cap.txt", T)\\
\}\\
\\
\noindent crypto.hist.prc <- function (date)\\
\{\\
\indent require(XML)\\
\indent require(httr)\\
\\
\indent x <- read.delim("crypto.cap.txt", header = T)\\
\indent x <- as.matrix(x)\\
\indent univ <- x[, 11]\\
\indent for(i in 1:length(univ))\\
\indent \{\\
\indent \indent url <- paste("https://coinmarketcap.com/currencies/",\\
\indent \indent \indent univ[i],\\
\indent \indent \indent "/historical-data/?start=20000101\&end=", date, sep = "")\\
\indent \indent x <- shared.get.webpage(url)\\
\indent \indent x <- shared.parse.html(x, keyword = "table")\\
\indent \indent n <- length(x)/7\\
\indent \indent if(n < 1 | trunc(n) != n)\\
\indent \indent \{\\
\indent \indent \indent write(univ[i], "crypto.bad.txt", append = T)\\
\indent \indent \indent next\\
\indent \indent \}\\
\indent \indent x <- matrix(x, length(x)/7, 7, byrow = T)\\
\indent \indent x[1, ] <- c("Date", "Open", "High", "Low",\\
\indent \indent \indent "Close", "Volume", "MktCap")\\
\indent \indent file <- paste("CryptoHistData/", univ[i], ".txt", sep = "")\\
\indent \indent shared.write.table(x, file, T)\\
\indent \}\\
\}\\
\\
\noindent crypto.prc.files <- function ()\\
\{\\
\indent match.univ <- function (univ1, univ2)\\
\indent \{\\
\indent \indent good <- match(univ1, univ2, nomatch = 0)\\
\indent \indent univ <- univ2[good]\\
\indent \indent return(univ)\\
\indent \}\\
\\
\indent read.file <- function(file, header = T)\\
\indent \{\\
\indent \indent x <- read.delim(file, header = header)\\
\indent \indent x <- as.matrix(x)\\
\indent \indent x <- gsub(",", "", x)\\
\indent \indent return(x)\\
\indent \}\\
\\
\indent x <- read.file("crypto.cap.txt")\\
\indent univ <- x[, 11]\\
\indent mnbl <- x[, 12]\\
\indent name <- x[, 2]\\
\indent bad <- readLines("crypto.bad.txt")\\
\indent take <- is.na(match(univ, bad))\\
\indent mnbl <- mnbl[take]\\
\indent name <- name[take]\\
\indent mnbl[mnbl == "Y"] <- 1\\
\indent mnbl[mnbl == "N"] <- 0\\
\indent univ <- univ[take]\\
\indent shared.write.table(mnbl, file = "cr.mnbl.txt", T)\\
\indent shared.write.table(name, file = "cr.name.txt", T)\\
\indent n <- length(univ)\\
\indent for(i in 1:n)\\
\indent \{\\
\indent \indent x <- read.file(paste("CryptoHistData/", univ[i],\\
\indent \indent \indent ".txt", sep = ""))\\
\indent \indent if(i == 1)\\
\indent \indent \{\\
\indent \indent \indent dates <- x[, "Date"]\\
\indent \indent \indent d <- length(dates)\\
\indent \indent \indent prc <- matrix(NA, n, d)\\
\indent \indent \indent cap <- matrix(NA, n, d)\\
\indent \indent \indent high <- matrix(NA, n, d)\\
\indent \indent \indent low <- matrix(NA, n, d)\\
\indent \indent \indent vol <- matrix(NA, n, d)\\
\indent \indent \indent open <- matrix(NA, n, d)\\
\indent \indent \indent dimnames(prc)[[2]] <- dates\\
\indent \indent \indent dimnames(cap)[[2]] <- dates\\
\indent \indent \indent dimnames(high)[[2]] <- dates\\
\indent \indent \indent dimnames(low)[[2]] <- dates\\
\indent \indent \indent dimnames(vol)[[2]] <- dates\\
\indent \indent \indent dimnames(open)[[2]] <- dates\\
\indent \indent \indent prc[1, ] <- x[1:d, "Close"]\\
\indent \indent \indent cap[1, ] <- x[1:d, "MktCap"]\\
\indent \indent \indent high[1, ] <- x[1:d, "High"]\\
\indent \indent \indent low[1, ] <- x[1:d, "Low"]\\
\indent \indent \indent vol[1, ] <- x[1:d, "Volume"]\\
\indent \indent \indent open[1, ] <- x[1:d, "Open"]\\
\indent \indent \}\\
\indent \indent else\\
\indent \indent \{\\
\indent \indent \indent dates1 <- x[, "Date"]\\
\indent \indent \indent dates1 <- match.univ(dates, dates1)\\
\indent \indent \indent prc[i, dates1] <- x[1:length(dates1), "Close"]\\
\indent \indent \indent cap[i, dates1] <- x[1:length(dates1), "MktCap"]\\
\indent \indent \indent high[i, dates1] <- x[1:length(dates1), "High"]\\
\indent \indent \indent low[i, dates1] <- x[1:length(dates1), "Low"]\\
\indent \indent \indent vol[i, dates1] <- x[1:length(dates1), "Volume"]\\
\indent \indent \indent open[i, dates1] <- x[1:length(dates1), "Open"]\\
\indent \indent \}\\
\indent \}\\
\indent mode(prc) <- "numeric"\\
\indent mode(cap) <- "numeric"\\
\indent mode(high) <- "numeric"\\
\indent mode(low) <- "numeric"\\
\indent mode(vol) <- "numeric"\\
\indent mode(open) <- "numeric"\\
\indent shared.write.table(prc, file = "cr.prc.txt", T)\\
\indent shared.write.table(cap, file = "cr.cap.txt", T)\\
\indent shared.write.table(high, file = "cr.high.txt", T)\\
\indent shared.write.table(low, file = "cr.low.txt", T)\\
\indent shared.write.table(vol, file = "cr.vol.txt", T)\\
\indent shared.write.table(open, file = "cr.open.txt", T)\\
\}\\
\\
\noindent crypto.prc <- function (days = 365, back = 0,\\
\indent lookback = days, d.r = 20, d.v = 20, d.i = 20)\\
\{\\
\indent calc.ix <- function(z, days)\\
\indent \{\\
\indent \indent ix <- colSums(z[, 1:days])\\
\indent \indent ix <- ix[days:1]\\
\indent \indent ix <- ix / ix[1]\\
\indent \indent return(ix)\\
\indent \}\\
\\
\indent read.prc <- function(file, header = F, make.numeric = T)\\
\indent \{\\
\indent \indent x <- read.delim(file, header = header)\\
\indent \indent x <- as.matrix(x)\\
\indent \indent if(make.numeric)\\
\indent \indent \indent mode(x) <- "numeric"\\
\indent \indent return(x)\\
\indent \}\\
\\
\indent calc.mv.avg <- function(x, days, d.r)\\
\indent \{\\
\indent \indent if(d.r == 1)\\
\indent \indent \indent return(x[, 1:days])\\
\indent \indent y <- matrix(0, nrow(x), days)\\
\indent \indent for(i in 1:days)\\
\indent \indent \indent y[, i] <- rowMeans(x[, i:(i + d.r - 1)], na.rm = T)\\
\\
\indent \indent return(y)\\
\indent \}\\
\\
\indent prc <- read.prc("cr.prc.txt")\\
\indent cap <- read.prc("cr.cap.txt")\\
\indent high <- read.prc("cr.high.txt")\\
\indent low <- read.prc("cr.low.txt")\\
\indent vol <- read.prc("cr.vol.txt")\\
\indent open <- read.prc("cr.open.txt")\\
\indent mnbl <- read.prc("cr.mnbl.txt")\\
\indent name <- read.prc("cr.name.txt", make.numeric = F)\\
\indent d <- days + d.r + 1\\
\indent prc <- prc[, 1:d]\\
\indent cap <- cap[, 1:d]\\
\indent high <- high[, 1:d]\\
\indent low <- low[, 1:d]\\
\indent vol <- vol[, 1:d]\\
\indent open <- open[, 1:d]\\
\\
\indent take <- rowSums(is.na(prc)) == 0 \& rowSums(is.na(cap)) == 0 \&\\
\indent \indent rowSums(is.na(high)) == 0 \& rowSums(is.na(low)) == 0 \&\\
\indent \indent rowSums(is.na(vol)) == 0 \& rowSums(is.na(open)) == 0 \&\\
\indent \indent rowSums(vol == 0) == 0\\
\\
\indent ret <- log(prc[take, -d] / prc[take, -1])\\
\indent prc <- prc[take, -1]\\
\indent cap <- cap[take, -1]\\
\indent high <- high[take, -1]\\
\indent low <- low[take, -1]\\
\indent vol <- vol[take, -1]\\
\indent open <- open[take, -1]\\
\indent mnbl <- mnbl[take, 1]\\
\indent name <- name[take, 1]\\
\indent if(back > 0)\\
\indent \{\\
\indent \indent ret <- ret[, (back + 1):ncol(ret)]\\
\indent \indent prc <- prc[, (back + 1):ncol(prc)]\\
\indent \indent cap <- cap[, (back + 1):ncol(cap)]\\
\indent \indent high <- high[, (back + 1):ncol(high)]\\
\indent \indent low <- low[, (back + 1):ncol(low)]\\
\indent \indent vol <- vol[, (back + 1):ncol(vol)]\\
\indent \indent open <- open[, (back + 1):ncol(open)]\\
\indent \}\\
\indent days <- lookback\\
\indent av <- log(calc.mv.avg(vol, days, d.v))\\
\indent hlv <- (high - low)\^{}2 / prc\^{}2\\
\indent hlv <- 0.5 * log(calc.mv.avg(hlv, days, d.i))\\
\indent take <- rowSums(!is.finite(hlv)) == 0\\
\indent av <- av[take, ]\\
\indent hlv <- hlv[take, ]\\
\indent mom <- log(prc / open)[take, 1:days]\\
\indent mom1 <- log(prc / open)[take, 1:days + 1]\\
\indent mom2 <- log(prc / open)[take, 1:days + 2]\\
\indent mom3 <- log(prc / open)[take, 1:days + 3]\\
\indent mom4 <- log(prc / open)[take, 1:days + 4]\\
\indent size <- log(cap)[take, 1:days]\\
\indent ret <- ret[take, 1:days]\\
\indent mnbl <- mnbl[take]\\
\indent name <- name[take]\\
\indent for(i in 1:days)\\
\indent \{\\
\indent \indent flm <- cbind(size[, i], mom[, i], hlv[, i], av[, i])\\
\indent \indent if(i == 1)\\
\indent \indent \indent fac <- matrix(NA, ncol(flm) + 1, days)\\
\indent \indent reg <- lm(ret[, i] ~ flm)\\
\indent \indent fac[, i] <- coefficients(reg)\\
\indent \}\\
\indent t.stat <- sqrt(365) * rowMeans(fac) / apply(fac, 1, sd)\\
\indent t.stat <- round(t.stat, 2)\\
\\
\indent prc <- prc[take, ]\\
\indent cap <- cap[take, ]\\
\indent y <- prc[name == "Xaurum", ]\\
\indent prc[name == "Xaurum", y > 1] <-\\
\indent \indent prc[name == "Xaurum", y > 1] / 8000\\
\indent ix.cap <- calc.ix(cap, days)\\
\indent ix.prc <- calc.ix(prc, days)\\
\indent plot(1:length(ix.cap), ix.cap, type = "l",\\
\indent \indent col = "green", xlab = "days", ylab = "index value",\\
\indent \indent ylim = c(min(c(ix.cap, ix.prc)) - .5,\\
\indent \indent \indent max(c(ix.cap, ix.prc)) + .5))\\
\indent lines(1:length(ix.prc), ix.prc, col = "blue")\\
\\
\indent return(t.stat)\\
\}
}}

\section{DISCLAIMERS}\label{app.B}

{}Wherever the context so requires, the masculine gender includes the feminine and/or neuter, and the singular form includes the plural and {\em vice versa}. The author of this paper (``Author") and his affiliates including without limitation Quantigic$^\circledR$ Solutions LLC (``Author's Affiliates" or ``his Affiliates") make no implied or express warranties or any other representations whatsoever, including without limitation implied warranties of merchantability and fitness for a particular purpose, in connection with or with regard to the content of this paper including without limitation any code or algorithms contained herein (``Content").

{}The reader may use the Content solely at his/her/its own risk and the reader shall have no claims whatsoever against the Author or his Affiliates and the Author and his Affiliates shall have no liability whatsoever to the reader or any third party whatsoever for any loss, expense, opportunity cost, damages or any other adverse effects whatsoever relating to or arising from the use of the Content by the reader including without any limitation whatsoever: any direct, indirect, incidental, special, consequential or any other damages incurred by the reader, however caused and under any theory of liability; any loss of profit (whether incurred directly or indirectly), any loss of goodwill or reputation, any loss of data suffered, cost of procurement of substitute goods or services, or any other tangible or intangible loss; any reliance placed by the reader on the completeness, accuracy or existence of the Content or any other effect of using the Content; and any and all other adversities or negative effects the reader might encounter in using the Content irrespective of whether the Author or his Affiliates is or are or should have been aware of such adversities or negative effects.

{}The R code included in Appendix \ref{app.A} hereof is part of the copyrighted R code of Quantigic$^\circledR$ Solutions LLC and is provided herein with the express permission of Quantigic$^\circledR$ Solutions LLC. The copyright owner retains all rights, title and interest in and to its copyrighted source code included in Appendix \ref{app.A} hereof and any and all copyrights therefor.


\newpage


\begin{table}[ht]
\caption{Results for regressions (\ref{reg}) with int (intercept) plus 4 factors cap, mom, hlv and vol defined in Section \ref{sec2} (along with parameters $d_{vol}$ and $d_{hlv}$). The universe is based on cryptoassets with historical data with non-NA close, high, low, open, volume and market cap, and nonzero volume for a lookback of $386 = 365 + 20 + 1$ days from August 18, 2018 (inclusive). This universe consists of 362 cryptoassets. The regressions are run over 365 days starting with the most recent date in the time series and going back in time. The t-statistic (t-stat) is annualized, by multiplying the daily t-statistic by $\sqrt{365}$. The same notations are used in the remainder of the tables below.} 
\begin{tabular}{l l l l l l l} 
\\
\hline\hline 
$d_{vol}$ & $d_{hlv}$ & t-stat:int & t-stat:cap & t-stat:mom & t-stat:hlv & t-stat:vol\\
20	&	20	&	2.66	&	-3.68	&	-34.2	&	-2.48	&	1.22	\\
20	&	15	&	2.59	&	-3.73	&	-34.17	&	-2.71	&	1.15	\\
20	&	10	&	2.54	&	-3.67	&	-34.18	&	-2.58	&	1.1	\\
20	&	5	&	2.51	&	-3.71	&	-34.19	&	-2.46	&	1.16	\\
15	&	20	&	2.59	&	-3.54	&	-34.19	&	-2.47	&	1.06	\\
15	&	15	&	2.56	&	-3.61	&	-34.16	&	-2.71	&	1.08	\\
15	&	10	&	2.52	&	-3.55	&	-34.17	&	-2.58	&	1.03	\\
15	&	5	&	2.47	&	-3.58	&	-34.18	&	-2.46	&	1.08	\\
10	&	20	&	2.53	&	-3.38	&	-34.17	&	-2.43	&	0.84	\\
10	&	15	&	2.51	&	-3.46	&	-34.15	&	-2.67	&	0.89	\\
10	&	10	&	2.52	&	-3.42	&	-34.18	&	-2.56	&	0.94	\\
10	&	5	&	2.49	&	-3.48	&	-34.19	&	-2.46	&	1	\\
5	&	20	&	2.36	&	-3.07	&	-34.14	&	-2.4	&	0.48	\\
5	&	15	&	2.34	&	-3.15	&	-34.12	&	-2.62	&	0.53	\\
5	&	10	&	2.36	&	-3.13	&	-34.15	&	-2.51	&	0.61	\\
5	&	5	&	2.42	&	-3.21	&	-34.19	&	-2.42	&	0.85	\\
3	&	20	&	2.23	&	-2.87	&	-34.06	&	-2.42	&	0.27	\\
3	&	15	&	2.21	&	-2.95	&	-34.04	&	-2.64	&	0.31	\\
3	&	10	&	2.22	&	-2.92	&	-34.08	&	-2.53	&	0.38	\\
3	&	5	&	2.27	&	-3	&	-34.12	&	-2.43	&	0.63	\\
1	&	20	&	3.47	&	-5.1	&	-33.98	&	-2.27	&	2.6	\\
1	&	15	&	3.44	&	-5.17	&	-33.96	&	-2.52	&	2.6	\\
1	&	10	&	3.43	&	-5.11	&	-34	&	-2.46	&	2.63	\\
1	&	5	&	3.5	&	-5.24	&	-34.06	&	-2.44	&	2.87	\\ [1ex] 
\hline 
\end{tabular}
\label{table1} 
\end{table}
\newpage
\begin{table}[ht]
\caption{Results for regressions (\ref{reg}) with int (intercept) plus 4 factors cap, mom, hlv and vol defined in Section \ref{sec2} (along with parameters $d_{vol}$ and $d_{hlv}$). The universe is based on cryptoassets with historical data with non-NA close, high, low, open, volume and market cap, and nonzero volume for a lookback of $751 = 2 \times 365 + 20 + 1$ days from August 18, 2018 (inclusive). This universe consists of 127 cryptoassets (after eliminating 2 cryptoassets with apparently ``artifact" stale prices; however, these stale prices occur prior to the most recent year in the time series, so these 2 cryptoassets are kept when the regressions are run only over said year, which is the case in some of the other tables). The regressions are run over $730 = 2 \times 365$ days starting with the most recent date in the time series and going back in time (i.e., over the 2 years in the 2-year time series).} 
\begin{tabular}{l l l l l l l} 
\\
\hline\hline 
$d_{vol}$ & $d_{hlv}$ & t-stat:int & t-stat:cap & t-stat:mom & t-stat:hlv & t-stat:vol\\
20	&	20	&	1.83	&	-1.45	&	-16.87	&	-2.72	&	-0.06	\\
20	&	15	&	1.81	&	-1.45	&	-16.84	&	-2.7	&	-0.07	\\
20	&	10	&	1.69	&	-1.48	&	-16.79	&	-3.23	&	-0.14	\\
20	&	5	&	1.54	&	-1.37	&	-16.75	&	-3.16	&	-0.26	\\
15	&	20	&	1.84	&	-1.45	&	-16.89	&	-2.71	&	-0.1	\\
15	&	15	&	1.86	&	-1.49	&	-16.85	&	-2.67	&	-0.01	\\
15	&	10	&	1.74	&	-1.53	&	-16.81	&	-3.19	&	-0.07	\\
15	&	5	&	1.58	&	-1.44	&	-16.76	&	-3.15	&	-0.19	\\
10	&	20	&	1.8	&	-1.39	&	-16.89	&	-2.69	&	-0.2	\\
10	&	15	&	1.83	&	-1.44	&	-16.85	&	-2.63	&	-0.09	\\
10	&	10	&	1.78	&	-1.55	&	-16.8	&	-3.11	&	0.03	\\
10	&	5	&	1.65	&	-1.49	&	-16.76	&	-3.09	&	-0.08	\\
5	&	20	&	1.68	&	-1.29	&	-16.85	&	-2.69	&	-0.38	\\
5	&	15	&	1.72	&	-1.36	&	-16.82	&	-2.63	&	-0.25	\\
5	&	10	&	1.68	&	-1.48	&	-16.77	&	-3.08	&	-0.11	\\
5	&	5	&	1.71	&	-1.61	&	-16.72	&	-3.04	&	0.12	\\
3	&	20	&	1.49	&	-1	&	-16.77	&	-2.63	&	-0.64	\\
3	&	15	&	1.53	&	-1.06	&	-16.73	&	-2.54	&	-0.54	\\
3	&	10	&	1.48	&	-1.16	&	-16.69	&	-2.98	&	-0.42	\\
3	&	5	&	1.51	&	-1.29	&	-16.66	&	-2.93	&	-0.19	\\
1	&	20	&	2.29	&	-2.26	&	-16.72	&	-2.8	&	0.75	\\
1	&	15	&	2.31	&	-2.28	&	-16.69	&	-2.73	&	0.84	\\
1	&	10	&	2.26	&	-2.37	&	-16.66	&	-3.18	&	0.94	\\
1	&	5	&	2.3	&	-2.55	&	-16.63	&	-3.2	&	1.16	\\ [1ex] 
\hline 
\end{tabular}
\label{table2} 
\end{table}
\newpage
\begin{table}[ht]
\caption{Results for regressions (\ref{reg}) with int (intercept) plus 4 factors cap, mom, hlv and vol defined in Section \ref{sec2} (along with parameters $d_{vol}$ and $d_{hlv}$). The universe is based on cryptoassets with historical data with non-NA close, high, low, open, volume and market cap, and nonzero volume for a lookback of $751 = 2 \times 365 + 20 + 1$ days from August 18, 2018 (inclusive). This universe consists of 129 cryptoassets. The regressions are run over $365$ days starting with the most recent date in the time series and going back in time (i.e., over the 1st year in the 2-year time series).} 
\begin{tabular}{l l l l l l l} 
\\
\hline\hline 
$d_{vol}$ & $d_{hlv}$ & t-stat:int & t-stat:cap & t-stat:mom & t-stat:hlv & t-stat:vol\\
20	&	20	&	-0.39	&	0.18	&	-17.17	&	-1.01	&	-0.67	\\
20	&	15	&	-0.36	&	0.19	&	-17.04	&	-0.96	&	-0.69	\\
20	&	10	&	-0.35	&	0.01	&	-17	&	-1.55	&	-0.66	\\
20	&	5	&	-0.35	&	0.04	&	-16.99	&	-1.6	&	-0.71	\\
15	&	20	&	-0.42	&	0.21	&	-17.16	&	-0.99	&	-0.71	\\
15	&	15	&	-0.37	&	0.2	&	-17.06	&	-0.89	&	-0.69	\\
15	&	10	&	-0.34	&	0.01	&	-17.01	&	-1.49	&	-0.64	\\
15	&	5	&	-0.35	&	0.03	&	-17	&	-1.56	&	-0.68	\\
10	&	20	&	-0.58	&	0.48	&	-17.15	&	-0.98	&	-1.03	\\
10	&	15	&	-0.53	&	0.48	&	-17.05	&	-0.87	&	-1	\\
10	&	10	&	-0.47	&	0.22	&	-17.05	&	-1.41	&	-0.87	\\
10	&	5	&	-0.47	&	0.24	&	-17.03	&	-1.51	&	-0.91	\\
5	&	20	&	-0.82	&	0.79	&	-17.08	&	-1	&	-1.38	\\
5	&	15	&	-0.76	&	0.76	&	-16.98	&	-0.9	&	-1.33	\\
5	&	10	&	-0.71	&	0.5	&	-16.99	&	-1.43	&	-1.18	\\
5	&	5	&	-0.67	&	0.47	&	-17.07	&	-1.39	&	-1.09	\\
3	&	20	&	-1.04	&	1.15	&	-16.98	&	-1.02	&	-1.79	\\
3	&	15	&	-0.99	&	1.13	&	-16.88	&	-0.91	&	-1.74	\\
3	&	10	&	-0.95	&	0.89	&	-16.89	&	-1.42	&	-1.6	\\
3	&	5	&	-0.9	&	0.85	&	-16.98	&	-1.34	&	-1.49	\\
1	&	20	&	-0.28	&	0.09	&	-16.69	&	-1.12	&	-0.69	\\
1	&	15	&	-0.21	&	0.05	&	-16.59	&	-1.03	&	-0.64	\\
1	&	10	&	-0.18	&	-0.17	&	-16.6	&	-1.57	&	-0.53	\\
1	&	5	&	-0.13	&	-0.24	&	-16.67	&	-1.57	&	-0.43	\\  [1ex] 
\hline 
\end{tabular}
\label{table3} 
\end{table}

\newpage
\begin{table}[ht]
\caption{Results for regressions (\ref{reg}) with int (intercept) plus 4 factors cap, mom, hlv and vol defined in Section \ref{sec2} (along with parameters $d_{vol}$ and $d_{hlv}$). The universe is based on cryptoassets with historical data with non-NA close, high, low, open, volume and market cap, and nonzero volume for a lookback of $751 = 2 \times 365 + 20 + 1$ days from August 18, 2018 (inclusive). This universe consists of 127 cryptoassets (after eliminating 2 cryptoassets with apparently ``artifact" stale prices). The regressions are run over $365$ days by skipping 365 most recent days, starting with the 366th day in the time series and going back in time (i.e., over the 2nd year in the 2-year time series).} 
\begin{tabular}{l l l l l l l} 
\\
\hline\hline 
$d_{vol}$ & $d_{hlv}$ & t-stat:int & t-stat:cap & t-stat:mom & t-stat:hlv & t-stat:vol\\
20	&	20	&	4.39	&	-2.43	&	-15.37	&	-2.84	&	0.37	\\
20	&	15	&	4.3	&	-2.43	&	-15.43	&	-2.86	&	0.38	\\
20	&	10	&	4.22	&	-2.36	&	-15.4	&	-2.96	&	0.19	\\
20	&	5	&	4.04	&	-2.28	&	-15.33	&	-2.9	&	0.09	\\
15	&	20	&	4.43	&	-2.42	&	-15.38	&	-2.83	&	0.28	\\
15	&	15	&	4.34	&	-2.44	&	-15.44	&	-2.83	&	0.41	\\
15	&	10	&	4.28	&	-2.38	&	-15.41	&	-2.92	&	0.23	\\
15	&	5	&	4.12	&	-2.33	&	-15.34	&	-2.89	&	0.12	\\
10	&	20	&	4.54	&	-2.58	&	-15.38	&	-2.87	&	0.4	\\
10	&	15	&	4.44	&	-2.61	&	-15.44	&	-2.88	&	0.54	\\
10	&	10	&	4.36	&	-2.53	&	-15.41	&	-2.88	&	0.53	\\
10	&	5	&	4.27	&	-2.52	&	-15.34	&	-2.86	&	0.45	\\
5	&	20	&	4.62	&	-2.81	&	-15.39	&	-2.95	&	0.49	\\
5	&	15	&	4.54	&	-2.83	&	-15.46	&	-2.94	&	0.64	\\
5	&	10	&	4.48	&	-2.77	&	-15.42	&	-2.93	&	0.65	\\
5	&	5	&	4.5	&	-2.86	&	-15.34	&	-2.89	&	0.88	\\
3	&	20	&	4.43	&	-2.64	&	-15.33	&	-2.97	&	0.38	\\
3	&	15	&	4.35	&	-2.65	&	-15.4	&	-2.93	&	0.51	\\
3	&	10	&	4.3	&	-2.59	&	-15.36	&	-2.9	&	0.51	\\
3	&	5	&	4.32	&	-2.68	&	-15.29	&	-2.87	&	0.73	\\
1	&	20	&	5.38	&	-4	&	-15.36	&	-3.03	&	2.07	\\
1	&	15	&	5.26	&	-3.94	&	-15.44	&	-2.97	&	2.15	\\
1	&	10	&	5.22	&	-3.91	&	-15.41	&	-2.98	&	2.16	\\
1	&	5	&	5.31	&	-4.11	&	-15.36	&	-3.02	&	2.39	\\  [1ex] 
\hline 
\end{tabular}
\label{table4} 
\end{table}

\newpage
\begin{table}[ht]
\caption{Results for regressions (\ref{reg}) with int (intercept) plus 4 factors cap, mom, hlv and vol defined in Section \ref{sec2} (along with parameters $d_{vol}$ and $d_{hlv}$). The universe is based on cryptoassets with historical data with non-NA close, high, low, open, volume and market cap, and nonzero volume for a lookback of $1116 = 3 \times 365 + 20 + 1$ days from August 18, 2018 (inclusive). This universe consists of 64 cryptoassets (after eliminating 2 cryptoassets with apparently ``artifact" stale prices). The regressions are run over $1095 = 3 \times 365$ days starting with the most recent date in the time series and going back in time (i.e., over the 3 years in the 3-year time series).} 
\begin{tabular}{l l l l l l l} 
\\
\hline\hline 
$d_{vol}$ & $d_{hlv}$ & t-stat:int & t-stat:cap & t-stat:mom & t-stat:hlv & t-stat:vol\\
20	&	20	&	1.38	&	-1.02	&	-12.64	&	-1.87	&	-0.15	\\
20	&	15	&	1.36	&	-1.07	&	-12.64	&	-1.96	&	-0.15	\\
20	&	10	&	1.32	&	-1.05	&	-12.62	&	-2.17	&	-0.21	\\
20	&	5	&	1.26	&	-1.12	&	-12.58	&	-2.31	&	-0.24	\\
15	&	20	&	1.44	&	-1.12	&	-12.68	&	-1.89	&	-0.08	\\
15	&	15	&	1.46	&	-1.2	&	-12.67	&	-1.94	&	-0.01	\\
15	&	10	&	1.43	&	-1.2	&	-12.66	&	-2.15	&	-0.05	\\
15	&	5	&	1.38	&	-1.3	&	-12.62	&	-2.3	&	-0.06	\\
10	&	20	&	1.33	&	-0.92	&	-12.67	&	-1.83	&	-0.28	\\
10	&	15	&	1.35	&	-1	&	-12.65	&	-1.86	&	-0.21	\\
10	&	10	&	1.37	&	-1.03	&	-12.63	&	-2.04	&	-0.18	\\
10	&	5	&	1.34	&	-1.19	&	-12.59	&	-2.26	&	-0.16	\\
5	&	20	&	1.19	&	-0.75	&	-12.75	&	-1.82	&	-0.44	\\
5	&	15	&	1.21	&	-0.82	&	-12.72	&	-1.83	&	-0.37	\\
5	&	10	&	1.23	&	-0.87	&	-12.68	&	-2	&	-0.33	\\
5	&	5	&	1.31	&	-1.11	&	-12.61	&	-2.12	&	-0.14	\\
3	&	20	&	1.09	&	-0.56	&	-12.74	&	-1.73	&	-0.6	\\
3	&	15	&	1.1	&	-0.61	&	-12.7	&	-1.71	&	-0.55	\\
3	&	10	&	1.12	&	-0.65	&	-12.67	&	-1.87	&	-0.51	\\
3	&	5	&	1.2	&	-0.89	&	-12.6	&	-1.99	&	-0.32	\\
1	&	20	&	1.78	&	-1.72	&	-12.63	&	-1.94	&	0.57	\\
1	&	15	&	1.79	&	-1.8	&	-12.62	&	-1.92	&	0.64	\\
1	&	10	&	1.8	&	-1.83	&	-12.6	&	-2.11	&	0.69	\\
1	&	5	&	1.86	&	-2.04	&	-12.55	&	-2.22	&	0.84	\\ [1ex] 
\hline 
\end{tabular}
\label{table5} 
\end{table}
\newpage
\begin{table}[ht]
\caption{Results for regressions (\ref{reg}) with int (intercept) plus 4 factors cap, mom, hlv and vol defined in Section \ref{sec2} (along with parameters $d_{vol}$ and $d_{hlv}$). The universe is based on cryptoassets with historical data with non-NA close, high, low, open, volume and market cap, and nonzero volume for a lookback of $1116 = 3 \times 365 + 20 + 1$ days from August 18, 2018 (inclusive). This universe consists of 64 cryptoassets (after eliminating 2 cryptoassets with apparently ``artifact" stale prices). The regressions are run over $730 = 2 \times 365$ days starting with the most recent date in the time series and going back in time (i.e., over the 1st and 2nd years in the 3-year time series).} 
\begin{tabular}{l l l l l l l} 
\\
\hline\hline 
$d_{vol}$ & $d_{hlv}$ & t-stat:int & t-stat:cap & t-stat:mom & t-stat:hlv & t-stat:vol\\
20	&	20	&	0.61	&	-0.38	&	-11.89	&	-2.2	&	-0.65	\\
20	&	15	&	0.59	&	-0.43	&	-11.86	&	-2.29	&	-0.64	\\
20	&	10	&	0.53	&	-0.39	&	-11.81	&	-2.63	&	-0.72	\\
20	&	5	&	0.44	&	-0.4	&	-11.71	&	-2.7	&	-0.8	\\
15	&	20	&	0.67	&	-0.47	&	-11.95	&	-2.22	&	-0.57	\\
15	&	15	&	0.7	&	-0.57	&	-11.89	&	-2.24	&	-0.49	\\
15	&	10	&	0.64	&	-0.55	&	-11.86	&	-2.58	&	-0.55	\\
15	&	5	&	0.55	&	-0.58	&	-11.76	&	-2.67	&	-0.62	\\
10	&	20	&	0.59	&	-0.31	&	-11.92	&	-2.17	&	-0.75	\\
10	&	15	&	0.63	&	-0.4	&	-11.86	&	-2.16	&	-0.67	\\
10	&	10	&	0.62	&	-0.44	&	-11.81	&	-2.44	&	-0.62	\\
10	&	5	&	0.53	&	-0.51	&	-11.71	&	-2.61	&	-0.68	\\
5	&	20	&	0.48	&	-0.17	&	-11.95	&	-2.17	&	-0.87	\\
5	&	15	&	0.51	&	-0.25	&	-11.88	&	-2.14	&	-0.79	\\
5	&	10	&	0.51	&	-0.31	&	-11.81	&	-2.42	&	-0.73	\\
5	&	5	&	0.55	&	-0.5	&	-11.69	&	-2.43	&	-0.57	\\
3	&	20	&	0.32	&	0.11	&	-11.92	&	-2.08	&	-1.15	\\
3	&	15	&	0.35	&	0.07	&	-11.84	&	-2	&	-1.1	\\
3	&	10	&	0.35	&	0.02	&	-11.78	&	-2.27	&	-1.05	\\
3	&	5	&	0.39	&	-0.16	&	-11.66	&	-2.27	&	-0.89	\\
1	&	20	&	0.88	&	-0.86	&	-11.7	&	-2.31	&	-0.17	\\
1	&	15	&	0.92	&	-0.95	&	-11.64	&	-2.25	&	-0.08	\\
1	&	10	&	0.91	&	-0.99	&	-11.59	&	-2.53	&	-0.04	\\
1	&	5	&	0.93	&	-1.13	&	-11.48	&	-2.5	&	0.09	\\ [1ex] 
\hline 
\end{tabular}
\label{table6} 
\end{table}
\newpage
\begin{table}[ht]
\caption{Results for regressions (\ref{reg}) with int (intercept) plus 4 factors cap, mom, hlv and vol defined in Section \ref{sec2} (along with parameters $d_{vol}$ and $d_{hlv}$). The universe is based on cryptoassets with historical data with non-NA close, high, low, open, volume and market cap, and nonzero volume for a lookback of $1116 = 3 \times 365 + 20 + 1$ days from August 18, 2018 (inclusive). This universe consists of 64 cryptoassets (after eliminating 2 cryptoassets with apparently ``artifact" stale prices). The regressions are run over $730 = 2 \times 365$ days by skipping the most recent 365 days, starting with the 366th day in the time series and going back in time (i.e., over the 2nd and 3rd years in the 3-year time series).} 
\begin{tabular}{l l l l l l l} 
\\
\hline\hline 
$d_{vol}$ & $d_{hlv}$ & t-stat:int & t-stat:cap & t-stat:mom & t-stat:hlv & t-stat:vol\\
20	&	20	&	3.16	&	-2.06	&	-12.97	&	-1.7	&	0.57	\\
20	&	15	&	3.14	&	-2.08	&	-13	&	-1.69	&	0.58	\\
20	&	10	&	3.09	&	-2.03	&	-12.94	&	-1.73	&	0.49	\\
20	&	5	&	3.09	&	-2.24	&	-12.94	&	-1.98	&	0.53	\\
15	&	20	&	3.27	&	-2.23	&	-13.03	&	-1.76	&	0.7	\\
15	&	15	&	3.26	&	-2.26	&	-13.05	&	-1.74	&	0.75	\\
15	&	10	&	3.22	&	-2.22	&	-13	&	-1.76	&	0.68	\\
15	&	5	&	3.23	&	-2.46	&	-13	&	-2.02	&	0.73	\\
10	&	20	&	3.13	&	-2.06	&	-13.06	&	-1.72	&	0.53	\\
10	&	15	&	3.13	&	-2.08	&	-13.06	&	-1.69	&	0.57	\\
10	&	10	&	3.11	&	-2.04	&	-13	&	-1.68	&	0.54	\\
10	&	5	&	3.19	&	-2.34	&	-13	&	-1.99	&	0.64	\\
5	&	20	&	2.92	&	-1.87	&	-13.18	&	-1.69	&	0.38	\\
5	&	15	&	2.92	&	-1.9	&	-13.19	&	-1.64	&	0.42	\\
5	&	10	&	2.9	&	-1.85	&	-13.12	&	-1.59	&	0.39	\\
5	&	5	&	3.06	&	-2.18	&	-13.06	&	-1.89	&	0.62	\\
3	&	20	&	2.87	&	-1.82	&	-13.22	&	-1.61	&	0.36	\\
3	&	15	&	2.86	&	-1.84	&	-13.23	&	-1.56	&	0.4	\\
3	&	10	&	2.84	&	-1.78	&	-13.15	&	-1.49	&	0.37	\\
3	&	5	&	2.98	&	-2.09	&	-13.1	&	-1.81	&	0.59	\\
1	&	20	&	3.88	&	-3.3	&	-13.26	&	-1.8	&	1.85	\\
1	&	15	&	3.88	&	-3.33	&	-13.28	&	-1.73	&	1.91	\\
1	&	10	&	3.85	&	-3.28	&	-13.22	&	-1.73	&	1.9	\\
1	&	5	&	3.97	&	-3.57	&	-13.18	&	-2.03	&	2.08	\\  [1ex] 
\hline 
\end{tabular}
\label{table7} 
\end{table}
\newpage
\begin{table}[ht]
\caption{Results for regressions (\ref{reg}) with int (intercept) plus 4 factors cap, mom, hlv and vol defined in Section \ref{sec2} (along with parameters $d_{vol}$ and $d_{hlv}$). The universe is based on cryptoassets with historical data with non-NA close, high, low, open, volume and market cap, and nonzero volume for a lookback of $1116 = 3 \times 365 + 20 + 1$ days from August 18, 2018 (inclusive). This universe consists of 66 cryptoassets. The regressions are run over $365$ days starting with the most recent date in the time series and going back in time (i.e., over the 1st year in the 3-year time series).} 
\begin{tabular}{l l l l l l l} 
\\
\hline\hline 
$d_{vol}$ & $d_{hlv}$ & t-stat:int & t-stat:cap & t-stat:mom & t-stat:hlv & t-stat:vol\\
20	&	20	&	-1.34	&	1.45	&	-10.55	&	-0.29	&	-1.74	\\
20	&	15	&	-1.29	&	1.44	&	-10.53	&	-0.33	&	-1.77	\\
20	&	10	&	-1.24	&	1.4	&	-10.59	&	-0.64	&	-1.81	\\
20	&	5	&	-1.19	&	1.37	&	-10.69	&	-0.73	&	-1.75	\\
15	&	20	&	-1.4	&	1.53	&	-10.52	&	-0.29	&	-1.82	\\
15	&	15	&	-1.32	&	1.45	&	-10.53	&	-0.27	&	-1.76	\\
15	&	10	&	-1.26	&	1.39	&	-10.59	&	-0.57	&	-1.78	\\
15	&	5	&	-1.2	&	1.35	&	-10.68	&	-0.68	&	-1.71	\\
10	&	20	&	-1.53	&	1.73	&	-10.48	&	-0.3	&	-2.05	\\
10	&	15	&	-1.44	&	1.63	&	-10.5	&	-0.27	&	-1.98	\\
10	&	10	&	-1.4	&	1.56	&	-10.58	&	-0.51	&	-1.98	\\
10	&	5	&	-1.35	&	1.55	&	-10.67	&	-0.64	&	-1.94	\\
5	&	20	&	-1.65	&	1.85	&	-10.46	&	-0.37	&	-2.2	\\
5	&	15	&	-1.54	&	1.73	&	-10.48	&	-0.33	&	-2.11	\\
5	&	10	&	-1.49	&	1.63	&	-10.57	&	-0.59	&	-2.07	\\
5	&	5	&	-1.49	&	1.62	&	-10.71	&	-0.53	&	-1.99	\\
3	&	20	&	-1.84	&	2.16	&	-10.4	&	-0.41	&	-2.55	\\
3	&	15	&	-1.74	&	2.06	&	-10.41	&	-0.36	&	-2.47	\\
3	&	10	&	-1.68	&	1.93	&	-10.5	&	-0.63	&	-2.42	\\
3	&	5	&	-1.65	&	1.88	&	-10.65	&	-0.53	&	-2.27	\\
1	&	20	&	-1.4	&	1.6	&	-10.09	&	-0.43	&	-1.9	\\
1	&	15	&	-1.28	&	1.47	&	-10.12	&	-0.4	&	-1.81	\\
1	&	10	&	-1.21	&	1.33	&	-10.2	&	-0.69	&	-1.75	\\
1	&	5	&	-1.14	&	1.25	&	-10.34	&	-0.61	&	-1.59	\\  [1ex] 
\hline 
\end{tabular}
\label{table8} 
\end{table}
\newpage
\begin{table}[ht]
\caption{Results for regressions (\ref{reg}) with int (intercept) plus 4 factors cap, mom, hlv and vol defined in Section \ref{sec2} (along with parameters $d_{vol}$ and $d_{hlv}$). The universe is based on cryptoassets with historical data with non-NA close, high, low, open, volume and market cap, and nonzero volume for a lookback of $1116 = 3 \times 365 + 20 + 1$ days from August 18, 2018 (inclusive). This universe consists of 64 cryptoassets (after eliminating 2 cryptoassets with apparently ``artifact" stale prices). The regressions are run over $365$ days by skipping the most recent 365 days, starting with the 366th day in the time series and going back in time (i.e., over the 2nd year in the 3-year time series).} 
\begin{tabular}{l l l l l l l} 
\\
\hline\hline 
$d_{vol}$ & $d_{hlv}$ & t-stat:int & t-stat:cap & t-stat:mom & t-stat:hlv & t-stat:vol\\
20	&	20	&	2.95	&	-1.71	&	-11.81	&	-2.25	&	0.25	\\
20	&	15	&	2.94	&	-1.72	&	-11.79	&	-2.15	&	0.27	\\
20	&	10	&	2.82	&	-1.62	&	-11.65	&	-2.29	&	0.14	\\
20	&	5	&	2.8	&	-1.83	&	-11.57	&	-2.5	&	0.15	\\
15	&	20	&	3.1	&	-1.92	&	-11.91	&	-2.34	&	0.42	\\
15	&	15	&	3.1	&	-1.94	&	-11.87	&	-2.2	&	0.48	\\
15	&	10	&	2.99	&	-1.86	&	-11.74	&	-2.32	&	0.36	\\
15	&	5	&	2.97	&	-2.07	&	-11.67	&	-2.52	&	0.37	\\
10	&	20	&	3.01	&	-1.8	&	-11.93	&	-2.32	&	0.31	\\
10	&	15	&	3.01	&	-1.8	&	-11.88	&	-2.16	&	0.35	\\
10	&	10	&	2.95	&	-1.75	&	-11.74	&	-2.2	&	0.31	\\
10	&	5	&	2.98	&	-2.02	&	-11.65	&	-2.48	&	0.34	\\
5	&	20	&	2.83	&	-1.67	&	-12.03	&	-2.3	&	0.25	\\
5	&	15	&	2.83	&	-1.68	&	-11.98	&	-2.12	&	0.29	\\
5	&	10	&	2.75	&	-1.6	&	-11.83	&	-2.11	&	0.23	\\
5	&	5	&	2.87	&	-1.9	&	-11.67	&	-2.33	&	0.44	\\
3	&	20	&	2.74	&	-1.55	&	-12.06	&	-2.24	&	0.13	\\
3	&	15	&	2.74	&	-1.56	&	-12.01	&	-2.05	&	0.18	\\
3	&	10	&	2.65	&	-1.45	&	-11.86	&	-2	&	0.09	\\
3	&	5	&	2.74	&	-1.72	&	-11.7	&	-2.22	&	0.29	\\
1	&	20	&	3.69	&	-2.9	&	-12.02	&	-2.46	&	1.48	\\
1	&	15	&	3.71	&	-2.93	&	-11.98	&	-2.25	&	1.56	\\
1	&	10	&	3.62	&	-2.87	&	-11.82	&	-2.27	&	1.51	\\
1	&	5	&	3.69	&	-3.1	&	-11.68	&	-2.41	&	1.66	\\  [1ex] 
\hline 
\end{tabular}
\label{table9} 
\end{table}
\newpage
\begin{table}[ht]
\caption{Results for regressions (\ref{reg}) with int (intercept) plus 4 factors cap, mom, hlv and vol defined in Section \ref{sec2} (along with parameters $d_{vol}$ and $d_{hlv}$). The universe is based on cryptoassets with historical data with non-NA close, high, low, open, volume and market cap, and nonzero volume for a lookback of $1116 = 3 \times 365 + 20 + 1$ days from August 18, 2018 (inclusive). This universe consists of 64 cryptoassets (after eliminating 2 cryptoassets with apparently ``artifact" stale prices). The regressions are run over $365$ days by skipping the most recent $730 = 2 \times 365$ days, starting with the 731st day in the time series and going back in time (i.e., over the 3rd year in the 3-year time series).} 
\begin{tabular}{l l l l l l l} 
\\
\hline\hline 
$d_{vol}$ & $d_{hlv}$ & t-stat:int & t-stat:cap & t-stat:mom & t-stat:hlv & t-stat:vol\\
20	&	20	&	3.53	&	-2.62	&	-14.15	&	-1.04	&	1.02	\\
20	&	15	&	3.48	&	-2.64	&	-14.23	&	-1.13	&	1	\\
20	&	10	&	3.53	&	-2.64	&	-14.26	&	-1.05	&	0.98	\\
20	&	5	&	3.54	&	-2.82	&	-14.35	&	-1.33	&	1.03	\\
15	&	20	&	3.56	&	-2.73	&	-14.17	&	-1.06	&	1.1	\\
15	&	15	&	3.55	&	-2.76	&	-14.25	&	-1.18	&	1.13	\\
15	&	10	&	3.61	&	-2.77	&	-14.28	&	-1.08	&	1.12	\\
15	&	5	&	3.65	&	-2.98	&	-14.37	&	-1.39	&	1.2	\\
10	&	20	&	3.38	&	-2.5	&	-14.2	&	-0.99	&	0.86	\\
10	&	15	&	3.36	&	-2.53	&	-14.27	&	-1.12	&	0.89	\\
10	&	10	&	3.41	&	-2.49	&	-14.28	&	-1.05	&	0.87	\\
10	&	5	&	3.52	&	-2.8	&	-14.38	&	-1.38	&	1.05	\\
5	&	20	&	3.13	&	-2.26	&	-14.35	&	-0.94	&	0.58	\\
5	&	15	&	3.11	&	-2.29	&	-14.43	&	-1.06	&	0.61	\\
5	&	10	&	3.18	&	-2.28	&	-14.44	&	-0.97	&	0.63	\\
5	&	5	&	3.39	&	-2.63	&	-14.49	&	-1.35	&	0.88	\\
3	&	20	&	3.11	&	-2.27	&	-14.4	&	-0.86	&	0.68	\\
3	&	15	&	3.08	&	-2.29	&	-14.47	&	-0.97	&	0.7	\\
3	&	10	&	3.14	&	-2.3	&	-14.48	&	-0.88	&	0.76	\\
3	&	5	&	3.35	&	-2.64	&	-14.53	&	-1.3	&	1	\\
1	&	20	&	4.24	&	-4.05	&	-14.53	&	-1	&	2.42	\\
1	&	15	&	4.19	&	-4.05	&	-14.61	&	-1.11	&	2.44	\\
1	&	10	&	4.24	&	-3.98	&	-14.66	&	-1.08	&	2.5	\\
1	&	5	&	4.43	&	-4.35	&	-14.76	&	-1.55	&	2.75	\\  [1ex] 
\hline 
\end{tabular}
\label{table10} 
\end{table}
\newpage
\begin{table}[ht]
\caption{Results for regressions (\ref{reg}) with int (intercept) plus 3 factors cap, mom and hlv  defined in Section \ref{sec2} (along with $d_{hlv}$). The period over which the regressions are run is the same as in the table shown in the first column below (where the corresponding cryptoasset universe is also defined).} 
\begin{tabular}{l l l l l l} 
\\
\hline\hline 
period & $d_{hlv}$ & t-stat:int & t-stat:cap & t-stat:mom & t-stat:hlv\\
Table \ref{table1}	&	20	&	2.07	&	-4.22	&	-34.11	&	-2.54	\\
Table \ref{table1}	&	15	&	2	&	-4.31	&	-34.08	&	-2.79	\\
Table \ref{table1}	&	10	&	1.97	&	-4.21	&	-34.09	&	-2.69	\\
Table \ref{table1}	&	5	&	1.88	&	-4.04	&	-34.09	&	-2.56	\\
Table \ref{table2}	&	20	&	2.42	&	-3.87	&	-16.76	&	-2.95	\\
Table \ref{table2}	&	15	&	2.4	&	-3.81	&	-16.72	&	-2.94	\\
Table \ref{table2}	&	10	&	2.24	&	-3.87	&	-16.68	&	-3.38	\\
Table \ref{table2}	&	5	&	2.09	&	-3.82	&	-16.63	&	-3.27	\\
Table \ref{table3}	&	20	&	-0.09	&	-0.98	&	-16.89	&	-1.1	\\
Table \ref{table3}	&	15	&	-0.05	&	-1.02	&	-16.76	&	-1.06	\\
Table \ref{table3}	&	10	&	-0.08	&	-1.38	&	-16.74	&	-1.66	\\
Table \ref{table3}	&	5	&	-0.06	&	-1.42	&	-16.76	&	-1.68	\\
Table \ref{table4}	&	20	&	5.68	&	-5.77	&	-15.34	&	-3.22	\\
Table \ref{table4}	&	15	&	5.6	&	-5.66	&	-15.41	&	-3.22	\\
Table \ref{table4}	&	10	&	5.38	&	-5.6	&	-15.37	&	-3.24	\\
Table \ref{table4}	&	5	&	5.05	&	-5.44	&	-15.3	&	-3.1	\\
[1ex] 
\hline 
\end{tabular}
\label{table11} 
\end{table}

\newpage
\begin{table}[ht]
\caption{Table \ref{table11} continued. Results for regressions (\ref{reg}) with int (intercept) plus 3 factors cap, mom and hlv  defined in Section \ref{sec2} (along with $d_{hlv}$). The period over which the regressions are run is the same as in the table shown in the first column below (where the corresponding cryptoasset universe is also defined).} 
\begin{tabular}{l l l l l l} 
\\
\hline\hline 
period & $d_{hlv}$ & t-stat:int & t-stat:cap & t-stat:mom & t-stat:hlv\\
Table \ref{table5}	&	20	&	1.9	&	-3.3	&	-12.57	&	-2.38	\\
Table \ref{table5}	&	15	&	1.86	&	-3.35	&	-12.57	&	-2.42	\\
Table \ref{table5}	&	10	&	1.86	&	-3.37	&	-12.5	&	-2.56	\\
Table \ref{table5}	&	5	&	1.75	&	-3.55	&	-12.44	&	-2.63	\\
Table \ref{table6}	&	20	&	1.27	&	-3.05	&	-11.81	&	-2.81	\\
Table \ref{table6}	&	15	&	1.24	&	-3.09	&	-11.76	&	-2.84	\\
Table \ref{table6}	&	10	&	1.21	&	-3.11	&	-11.64	&	-3.08	\\
Table \ref{table6}	&	5	&	1.08	&	-3.22	&	-11.52	&	-3.03	\\
Table \ref{table7}	&	20	&	3.71	&	-4.03	&	-13.02	&	-2.06	\\
Table \ref{table7}	&	15	&	3.68	&	-4.06	&	-13.04	&	-2.01	\\
Table \ref{table7}	&	10	&	3.62	&	-4.06	&	-12.95	&	-2.04	\\
Table \ref{table7}	&	5	&	3.5	&	-4.28	&	-12.93	&	-2.17	\\
Table \ref{table8}	&	20	&	-0.28	&	-0.31	&	-10.32	&	-0.42	\\
Table \ref{table8}	&	15	&	-0.19	&	-0.48	&	-10.32	&	-0.47	\\
Table \ref{table8}	&	10	&	-0.11	&	-0.75	&	-10.37	&	-0.75	\\
Table \ref{table8}	&	5	&	-0.13	&	-0.82	&	-10.44	&	-0.83	\\
Table \ref{table9}	&	20	&	3.72	&	-4.24	&	-11.92	&	-2.67	\\
Table \ref{table9}	&	15	&	3.7	&	-4.21	&	-11.88	&	-2.54	\\
Table \ref{table9}	&	10	&	3.55	&	-4.22	&	-11.67	&	-2.67	\\
Table \ref{table9}	&	5	&	3.39	&	-4.34	&	-11.57	&	-2.62	\\
Table \ref{table10}	&	20	&	3.82	&	-3.81	&	-14.12	&	-1.32	\\
Table \ref{table10}	&	15	&	3.77	&	-3.89	&	-14.2	&	-1.37	\\
Table \ref{table10}	&	10	&	3.82	&	-3.89	&	-14.24	&	-1.29	\\
Table \ref{table10}	&	5	&	3.8	&	-4.22	&	-14.32	&	-1.62	\\
[1ex] 
\hline 
\end{tabular}
\label{table12} 
\end{table}
\newpage
\begin{table}[ht]
\caption{Results for regressions (\ref{reg}) with int (intercept) plus 2 factors cap and mom defined in Section \ref{sec2}. The period over which the regressions are run is the same as in the table shown in the first column below (where the corresponding cryptoasset universe is also defined).} 
\begin{tabular}{l l l l} 
\\
\hline\hline 
period & t-stat:int & t-stat:cap & t-stat:mom\\
Table \ref{table1}	&	1.81	&	-3.01	&	-33.63	\\
Table \ref{table2}	&	2.2	&	-2.14	&	-16.5	\\
Table \ref{table3}	&	0.13	&	-0.48	&	-16.16	\\
Table \ref{table4}	&	4.77	&	-3.95	&	-15.11	\\
Table \ref{table5}	&	2	&	-1.98	&	-12.22	\\
Table \ref{table6}	&	1.45	&	-1.31	&	-11.44	\\
Table \ref{table7}	&	3.34	&	-2.92	&	-12.67	\\
Table \ref{table8}	&	0.24	&	-0.54	&	-9.8	\\
Table \ref{table9}	&	3.28	&	-2.5	&	-11.56	\\
Table \ref{table10}	&	3.47	&	-3.44	&	-13.78	\\
[1ex] 
\hline 
\end{tabular}
\label{table13} 
\end{table}
\newpage
\begin{table}[ht]
\caption{Results for regressions (\ref{reg}) {\em without} int (intercept) and with 3 factors cap, mom and hlv defined in Section \ref{sec2} (along with $d_{hlv}$) for $d_{hlv} = 5$. The period over which the regressions are run is the same as in the table shown in the first column below (where the corresponding cryptoasset universe is also defined).} 
\begin{tabular}{l l l l} 
\\
\hline\hline 
period & t-stat:cap & t-stat:mom & t-stat:hlv \\
Table \ref{table1}	&	-2.28	&	-33.23	&	-2.23	\\
Table \ref{table2}	&	-1.75	&	-16.33	&	-3.24	\\
Table \ref{table3}	&	-1.36	&	-16.47	&	-1.7	\\
Table \ref{table4}	&	-0.93	&	-15.03	&	-2.94	\\
Table \ref{table5}	&	-1.34	&	-12.24	&	-2.58	\\
Table \ref{table6}	&	-1.67	&	-11.29	&	-3	\\
Table \ref{table7}	&	-0.81	&	-12.75	&	-2.3	\\
Table \ref{table8}	&	-0.35	&	-10.08	&	-0.51	\\
Table \ref{table9}	&	-0.96	&	-11.38	&	-2.89	\\
Table \ref{table10}	&	-0.66	&	-14.15	&	-1.56	\\
[1ex] 
\hline 
\end{tabular}
\label{table14} 
\end{table}
\newpage
\begin{table}[ht]
\caption{Results for regressions (\ref{reg}) with int (intercept) plus 4 factors cap, mom, mom1 and hlv  defined in Section \ref{sec2} (along with $d_{hlv}$). The period over which the regressions are run is the same as in the table shown in the first column below (where the corresponding cryptoasset universe is also defined).} 
\begin{tabular}{l l l l l l l} 
\\
\hline\hline 
period & $d_{hlv}$ & t-stat:int & t-stat:cap & t-stat:mom & t-stat:mom1 & t-stat:hlv\\
Table \ref{table1}	&	20	&	1.69	&	-3.79	&	-35.89	&	-15.24	&	-2.32	\\
Table \ref{table1}	&	15	&	1.63	&	-3.85	&	-35.85	&	-15.2	&	-2.54	\\
Table \ref{table1}	&	10	&	1.62	&	-3.73	&	-35.86	&	-15.19	&	-2.37	\\
Table \ref{table1}	&	5	&	1.53	&	-3.48	&	-35.84	&	-15.13	&	-2.06	\\
Table \ref{table2}	&	20	&	2.21	&	-3.25	&	-18.05	&	-8.04	&	-2.47	\\
Table \ref{table2}	&	15	&	2.2	&	-3.18	&	-18	&	-8.02	&	-2.43	\\
Table \ref{table2}	&	10	&	2.03	&	-3.21	&	-17.96	&	-7.98	&	-2.83	\\
Table \ref{table2}	&	5	&	1.89	&	-3.08	&	-17.9	&	-7.79	&	-2.52	\\
Table \ref{table3}	&	20	&	-0.64	&	-0.32	&	-19.77	&	-9.46	&	-1.04	\\
Table \ref{table3}	&	15	&	-0.61	&	-0.28	&	-19.63	&	-9.41	&	-0.94	\\
Table \ref{table3}	&	10	&	-0.66	&	-0.64	&	-19.62	&	-9.43	&	-1.58	\\
Table \ref{table3}	&	5	&	-0.67	&	-0.54	&	-19.63	&	-9.26	&	-1.46	\\
Table \ref{table4}	&	20	&	5.49	&	-4.88	&	-16.54	&	-7.93	&	-2.33	\\
Table \ref{table4}	&	15	&	5.43	&	-4.83	&	-16.59	&	-7.91	&	-2.34	\\
Table \ref{table4}	&	10	&	5.2	&	-4.74	&	-16.54	&	-7.87	&	-2.31	\\
Table \ref{table4}	&	5	&	4.86	&	-4.53	&	-16.45	&	-7.69	&	-2.02	\\
[1ex] 
\hline 
\end{tabular}
\label{table15} 
\end{table}
\newpage
\begin{table}[ht]
\caption{Table \ref{table15} continued. Results for regressions (\ref{reg}) with int (intercept) plus 4 factors cap, mom, mom1 and hlv  defined in Section \ref{sec2} (along with $d_{hlv}$). The period over which the regressions are run is the same as in the table shown in the first column below (where the corresponding cryptoasset universe is also defined).} 
\begin{tabular}{l l l l l l l} 
\\
\hline\hline 
period & $d_{hlv}$ & t-stat:int & t-stat:cap & t-stat:mom & t-stat:mom1 & t-stat:hlv\\
Table \ref{table5}	&	20	&	1.86	&	-3.27	&	-13.4	&	-5.47	&	-2.37	\\
Table \ref{table5}	&	15	&	1.83	&	-3.32	&	-13.38	&	-5.45	&	-2.41	\\
Table \ref{table5}	&	10	&	1.78	&	-3.28	&	-13.28	&	-5.38	&	-2.54	\\
Table \ref{table5}	&	5	&	1.61	&	-3.39	&	-13.23	&	-5.2	&	-2.56	\\
Table \ref{table6}	&	20	&	1.2	&	-2.92	&	-12.44	&	-4.97	&	-2.73	\\
Table \ref{table6}	&	15	&	1.17	&	-2.95	&	-12.39	&	-4.95	&	-2.75	\\
Table \ref{table6}	&	10	&	1.1	&	-2.98	&	-12.25	&	-4.87	&	-3.03	\\
Table \ref{table6}	&	5	&	0.89	&	-2.96	&	-12.15	&	-4.65	&	-2.94	\\
Table \ref{table7}	&	20	&	3.73	&	-4.09	&	-14.01	&	-5.95	&	-2.13	\\
Table \ref{table7}	&	15	&	3.7	&	-4.11	&	-14.03	&	-5.94	&	-2.11	\\
Table \ref{table7}	&	10	&	3.57	&	-4.04	&	-13.91	&	-5.89	&	-2.14	\\
Table \ref{table7}	&	5	&	3.38	&	-4.19	&	-13.88	&	-5.76	&	-2.25	\\
Table \ref{table8}	&	20	&	-0.6	&	0.11	&	-11.67	&	-6.06	&	-0.44	\\
Table \ref{table8}	&	15	&	-0.49	&	0.01	&	-11.63	&	-6.03	&	-0.42	\\
Table \ref{table8}	&	10	&	-0.4	&	-0.25	&	-11.68	&	-5.96	&	-0.71	\\
Table \ref{table8}	&	5	&	-0.49	&	-0.18	&	-11.84	&	-6.02	&	-0.78	\\
Table \ref{table9}	&	20	&	3.71	&	-4.19	&	-12.68	&	-5.43	&	-2.69	\\
Table \ref{table9}	&	15	&	3.68	&	-4.15	&	-12.67	&	-5.42	&	-2.57	\\
Table \ref{table9}	&	10	&	3.46	&	-4.17	&	-12.46	&	-5.36	&	-2.81	\\
Table \ref{table9}	&	5	&	3.14	&	-4.14	&	-12.37	&	-5.2	&	-2.77	\\
Table \ref{table10}	&	20	&	3.91	&	-3.99	&	-15.39	&	-6.59	&	-1.46	\\
Table \ref{table10}	&	15	&	3.86	&	-4.08	&	-15.43	&	-6.57	&	-1.55	\\
Table \ref{table10}	&	10	&	3.87	&	-3.91	&	-15.4	&	-6.55	&	-1.3	\\
Table \ref{table10}	&	5	&	3.87	&	-4.27	&	-15.45	&	-6.42	&	-1.6	\\
[1ex] 
\hline 
\end{tabular}
\label{table16} 
\end{table}
\newpage
\begin{table}[ht]
\caption{Results for regressions (\ref{reg}) with int (intercept) plus 7 factors cap, mom, mom1, mom2, mom3, mom4 and hlv  defined in Section \ref{sec2} (along with $d_{hlv}$) for $d_{hlv} = 5$. The period over which the regressions are run is the same as in the table shown in the first column below (where the corresponding cryptoasset universe is also defined). The remaining columns show the corresponding t-stat.} 
\begin{tabular}{l l l l l l l l l} 
\\
\hline\hline 
period & int & cap & mom & mom1 & mom2 & mom3 & mom4 & hlv\\
Table \ref{table1}	&	0.95	&	-2.43	&	-37.16	&	-18.9	&	-10.61	&	-4.45	&	-3.97	&	-1.06	\\
Table \ref{table2}	&	1.47	&	-2.55	&	-18.26	&	-8.54	&	-4.63	&	-0.54	&	-0.65	&	-2.29	\\
Table \ref{table3}	&	-0.84	&	-0.09	&	-21.06	&	-10.45	&	-5.63	&	-2.03	&	-0.82	&	-1.08	\\
Table \ref{table4}	&	4.41	&	-3.61	&	-16.51	&	-8.14	&	-3.76	&	-0.9	&	-1.51	&	-1.56	\\
Table \ref{table5}	&	1.23	&	-2.83	&	-13.36	&	-5.59	&	-2.58	&	-0.8	&	-0.72	&	-2.38	\\
Table \ref{table6}	&	0.44	&	-2.56	&	-12.11	&	-4.96	&	-1.86	&	-0.56	&	-0.37	&	-2.93	\\
Table \ref{table7}	&	3.11	&	-3.28	&	-13.99	&	-6.18	&	-2.92	&	-1.41	&	-1.79	&	-1.49	\\
Table \ref{table8}	&	-0.66	&	-0.92	&	-13.17	&	-6.68	&	-2.15	&	-1.62	&	0.66	&	-1.76	\\
Table \ref{table9}	&	2.83	&	-3.2	&	-12.12	&	-5.53	&	-1.89	&	-1.59	&	-2.07	&	-1.97	\\
Table \ref{table10}	&	3.61	&	-3.37	&	-15.99	&	-6.99	&	-4.05	&	-1.24	&	-1.48	&	-0.83	\\
[1ex] 
\hline 
\end{tabular}
\label{table17} 
\end{table}
\newpage
\begin{table}[ht]
\caption{Results for regressions (\ref{reg}) with int (intercept) plus 4 factors cap, mom, hlv and mnbl defined in Section \ref{sec2} (along with $d_{hlv}$) for $d_{hlv} = 5$. The period over which the regressions are run is the same as in the table shown in the first column below (where the corresponding cryptoasset universe is also defined).} 
\begin{tabular}{l l l l l l} 
\\
\hline\hline 
period & t-stat:int & t-stat:cap & t-stat:mom & t-stat:hlv & t-stat:mnbl\\
Table \ref{table1}	&	1.88	&	-4.09	&	-34.19	&	-2.53	&	0.01	\\
Table \ref{table2}	&	2.25	&	-4.04	&	-16.67	&	-3.32	&	-0.88	\\
Table \ref{table3}	&	-0.13	&	-1.41	&	-16.74	&	-1.67	&	0.23	\\
Table \ref{table4}	&	5.17	&	-5.8	&	-15.37	&	-3.22	&	-1.5	\\
Table \ref{table5}	&	2.04	&	-3.68	&	-12.43	&	-2.61	&	-1.18	\\
Table \ref{table6}	&	1.27	&	-3.28	&	-11.53	&	-2.97	&	-0.74	\\
Table \ref{table7}	&	3.74	&	-4.39	&	-12.93	&	-2.21	&	-1.37	\\
Table \ref{table8}	&	-0.06	&	-0.79	&	-10.4	&	-0.76	&	-0.22	\\
Table \ref{table9}	&	3.41	&	-4.34	&	-11.63	&	-2.62	&	-0.77	\\
Table \ref{table10}	&	4.38	&	-4.46	&	-14.26	&	-1.72	&	-2.01	\\
[1ex] 
\hline 
\end{tabular}
\label{table18} 
\end{table}
%


\newpage\clearpage

\newpage\clearpage

\newpage\clearpage

\newpage\clearpage

\newpage\clearpage

\newpage\clearpage

\newpage\clearpage

\end{document}